\documentclass[review, english, sort&compress]{elsarticle}
\usepackage{amsmath}
\usepackage{amssymb}
\usepackage{amscd}
\usepackage{graphicx}
\usepackage{adjustbox}
\usepackage{pifont}
\usepackage{natbib} 
\usepackage{geometry}
\usepackage{hyperref}
\usepackage{amsfonts}
\usepackage{makeidx}
\usepackage{lmodern}
\usepackage[T1]{fontenc}
\usepackage{float}
\usepackage{multirow}
\usepackage{textcomp}
\usepackage[english]{babel}
\usepackage[flushleft]{threeparttable}
\usepackage{epstopdf}
\usepackage{bm} % define \pmb
\usepackage{xcolor}
\usepackage{pgf}
\usepackage{chngcntr}
\usepackage{pdfpages}

\makeatletter
\def\ps@pprintTitle{%
 \let\@oddhead\@empty
 \let\@evenhead\@empty
 \def\@oddfoot{}%
 \let\@evenfoot\@oddfoot}
\makeatother

\makeatletter
\renewcommand\appendix{\par
  \setcounter{section}{0}%
  \setcounter{subsection}{0}%
  \setcounter{equation}{0}%
  \setcounter{table}{0}%------------ << add
  \setcounter{figure}{0}%----------- << add
  \gdef\theequation{\@Alph\c@section.\arabic{equation}}%
  \gdef\thefigure{\@Alph\c@section.\arabic{figure}}%
  \gdef\thetable{\@Alph\c@section.\arabic{table}}%
  \gdef\thesection{\appendixname\@Alph\c@section}%
  \@addtoreset{equation}{section}%
  \@addtoreset{table}{section}%----- << add
  \@addtoreset{figure}{section}%---- << add
}
\makeatother

\newcommand{\AR}[1]{{\textcolor{red}{#1}}}
\newcommand{\comment}[1]{}

\biboptions{sort&compress}

\begin{document}
\sloppy
\begin{frontmatter}
\title{Interaction of $\langle a \rangle$ prismatic screw dislocations with the $\alpha-\beta$ interface side face in $\alpha-\beta$ Ti alloys}

\author{Ali Rida \corref{cor1}}
\ead{arida1@jhu.edu}
\author{Satish I. Rao \corref{}}
\author{Jaafar A. El-Awady \corref{cor1}}
\ead{jelawady@jhu.edu}

\cortext[cor1]{Corresponding authors}

\address{Department of Mechanical Engineering, Whiting School of Engineering, Johns Hopkins University, Baltimore, MD 21218, United States}

\begin{abstract}
Slip transmission across $\alpha-\beta$ interfaces is of great significance in understanding the strength of $\alpha-\beta$ Ti alloys for aerospace applications. Molecular statics (MS) and molecular dynamics (MD) simulations were conducted to investigate the mechanisms of slip transmission of $\langle a \rangle$ prismatic screw dislocations across the $\alpha-\beta$ interface side face. In these simulations, the $\alpha$ phase consisted of pure HCP Ti whereas the $\beta$ phase was modeled as Ti$_{60}$Nb$_{40}$ BCC random alloy using a Ti-Nb interatomic potential. Firstly, the misfit dislocations structure on the $\alpha-\beta$ interface side face has been characterized from MS simulations. This predicted dislocation structure is in good agreement with experimental observations in Ti-alloys. Secondly, MD simulations of the interaction of $[a_1]$, $[a_2]$, and $[a_3]$ prismatic screw dislocations with the interface side face in an $\alpha-\beta$  bi-crystal were performed at different temperatures. A distinct barrier of the interface side face to different types of dislocation transmission was found. The origin of this anisotropy in the slip transmission is due to the relative misalignment of the slip systems between the $\alpha$ and the $\beta$ phases from the Burgers orientation relationship. Finally, the mechanisms of slip transmission of each dislocation type were analyzed in a detailed atomistic description and compared to experimental observations.

\end{abstract}
\end{frontmatter}

%\tableofcontents

%\newpage

\section{Introduction}
Two phase $\alpha-\beta$ titanium alloys are widely used in many engineering applications due to their high specific strength, specific modulus, and toughness over a wide range of temperatures. Ti-alloys can be tailored to posses a combination of different favorable mechanical properties for specific applications by changing the thermomechanical processing route of these alloys. This can lead to a variety of microstructures that contain different distributions of $\alpha$ and $\beta$ phases \cite{lutjering2000microstructure}. Nevertheless, irrespective of the underlying microstructure in $\alpha-\beta$ Ti-alloys, the hexagonal close packed (HCP) $\alpha$ and the body centered cubic (BCC) $\beta$ phases are consistently observed to have a near classic Burgers orientation relationship (BOR) \cite{furuhara1996crystallography, ye2004tem, ye2006dislocation, zheng2018determination}. 

There are twelve variants of the BOR in which the $\lbrace{ 1 \bar{1} 0\rbrace}_{\beta} || \lbrace 0 0 0 1\rbrace_{\alpha}$, and the $\langle 1 1 1 \rangle_{\beta} || \langle 1 1 \bar{2}  0\rangle_{\alpha}$. This orientation relationship is important since it leads to the formation of low energy interfaces between the two phases \cite{furuhara1996crystallography, pond2003comparison}. The $\alpha$ precipitates usually have a lath or plate morphology elongated along a long axis (i.e, the invariant line direction). The broad face of the lath is called the habit plane with an irrational orientation close to the $\lbrace 1 1 \bar{1} \rbrace_{\beta}$ and the side facet is usually parallel to the basal plane for the $\alpha$ phase or the $\lbrace 1 \bar{1} 0 \rbrace_{\beta}$ \cite{ye2004tem, ye2006dislocation, zheng2018determination}. 

To accommodate the interface misfit between the two phases, the broad, side, and edge facets must contain interfacial defects, including misfit interfacial dislocations and structural ledges on the broad and edge facets \cite{ye2004tem, ye2006dislocation, zheng2018determination,zhang2021study}. A set of parallel misfit dislocations have been observed experimentally on the habit planes aligned along the invariant line direction \cite{furuhara1996crystallography, menon1986interfacial, ye2004tem, zheng2018determination, zhang2021study}. The Burgers vector for these misfit dislocations was $\langle c+a \rangle_{\alpha}$  \cite{ye2004tem, zheng2018determination}. On the other hand, misfit dislocations on other facets of the interface have received less attention. For instance, two sets of dislocations were observed on the side face having finely and coarsely dislocations spacing in Ti$-7.26 \ wt.\% \ $Cr alloy \cite{ye2006dislocation}. The Burgers vector of the fine spaced dislocations was estimated to be $1/2 \langle 1 1 1\rangle_{\beta} || 1/3 \langle 1 1\bar{2} 0 \rangle_{\alpha}$, whereas the Burgers vector of the coarse spaced dislocations was determined to be  $\langle 1 0 0\rangle_{\beta}$ in the $\beta$ phase or  $1/3 \langle 2 \bar{1} \bar{1} 0 \rangle_{\alpha}$ in the $\alpha$ phase \cite{ye2006dislocation}.

The interaction of $\langle a \rangle_{\alpha} = 1/3 \langle 1 1 
 \bar{2} 0\rangle_{\alpha}$ matrix dislocations with the $\alpha-\beta$ interfaces in these alloys is expected to control the properties of those alloys, including room temperature creep, yield strength, and strain hardening \cite{suri1999room, savage2004anisotropy}. A significant anisotropy of the strength of the $\alpha-\beta$ interface for different $\langle a \rangle_{\alpha}$ prismatic and basal slip matrix dislocations was measured experimentally\cite{suri1999room, savage2004anisotropy}. This behavior was attributed to the BOR between the $\alpha$ and the $\beta$ phases. However, a detailed description of the slip transmission mechanisms across the $\alpha-\beta$ interfaces is still largely missing. Zhao \textit{et al} adopted a microscopic phase field framework to simulating the transmission of a constant flux of discrete dislocations across multiple $\alpha-\beta$ interfaces in single colony Ti-alloys oriented for $\langle a \rangle_{\alpha}$ basal slip \cite{zhao2019slip}. Their results suggested that slip transmission is assisted by Shockley partials. They also suggested a new strengthening strategy by reducing the stacking fault energy of the $\alpha$ phase via alloying, which can lead to greater hindrance of the $\alpha-\beta$ interface to dislocation transfer, and hence, a higher hardening rate. Such studies have mainly focused on the slip transmission across the broad face of the $\alpha$ lath \cite{suri1999room, savage2004anisotropy, zhao2019slip}. However, prismatic $\langle a \rangle_{\alpha}$ screw dislocations can also interact with the interface side facet. To date, a detailed description of the slip transmission mechanisms of any $\langle a \rangle_{\alpha}$ type dislocation on the side facet is still missing, and whether the anisotropy of the strength of the interface to different $\langle a \rangle_{\alpha}$ dislocations is also present on this facet of the interface or not remains an open question. 

Accordingly, the aim of this study is to conduct molecular statics (MS) and dynamics (MD) simulations to investigate the slip transmission mechanisms of all three types of $\langle a \rangle_{\alpha}$ prismatic screw dislocations across the $\alpha-\beta$ interface side face in $\alpha-\beta$ Ti alloys. The objectives of this study are, first, to quantify if an anisotropy in the strength of the side face to $\langle a \rangle_{\alpha}$ screw dislocation type also exist, and second, to characterize the slip transmission mechanisms in a detailed atomistic description.

\section{Computational methods}
All MD simulations here were conducted using the open source large-scale Atomic/Molecular Massively Parallel Simulator (LAMMPS) \cite{thompson2022lammps}, with the Ti-Nb spline-like MEAM interatomic potential developed by Ehemann and Wilkins \cite{ehemann2017force}. This potential is referred to hereafter as the Ehemann potential. This potential was shown to predict the correct ground state of the $\langle a \rangle$ screw dislocation in pure $\alpha$ Ti \cite{rida2022characteristics} with a core dissociated on the pyramidal-I plane, which is in excellent agreement with ab-initio calculations \cite{clouet2015dislocation}. In addition,  Nb is a $\beta$ stabilizer for Ti and is continuously soluble in Ti \cite{moffat1988stable,ehemann2017force}. This is evident in Figure \ref{Fig:beta_phase_selection}(a), which shows the fraction of BCC atoms in a Ti$_{100-x}$Nb$_{x}$ random alloy for different atomic fractions of Nb$_x (\%)$ at 0 and 600K, as predicted by the Ehemann potential. For Nb concentrations higher than 35 at.$\%$, the fraction of BCC atoms in the simulation cell converge to 100\%, which indicates that the BCC phase becomes stable. Thus, this potential provides a suitable avenue for quantifying the deformation in dual-phase $\alpha-\beta$ Ti alloys. 

%*******************************
\subsection{Atomistic Simulations of the dual-phase $\alpha-\beta$ alloy}
In this work, a dual-phase $\alpha-\beta$ atomic system is constructed for the simulations. The  HCP $\alpha$ phase is modeled as pure Ti, while the BCC $\beta$ phase is modeled as Ti$_{60}$Nb$_{40}$ random alloy. The calculated lattice parameters using the Ehemann potential are $a_{\alpha} = 2.936 \AA$, $c_{\alpha} =  4.633 \AA$, for the $\alpha$ phase and $a_{\beta} = 3.274 \AA$ for the $\beta$ phase (see Figure \ref{Fig:beta_phase_selection}(b) in \ref{beta_phase}). This leads to a $\approx 3.5 \%$ difference between the Burgers vectors magnitude $\langle a \rangle_{\alpha}$ in the $\alpha$ phase and $1/2 \langle 111\rangle_{\beta}$ in the $\beta$ phase. This is in agreement with the experimentally reported range for many dual-phase Ti alloys including Ti-5553 ($\approx 2.6 \%$) \cite{zheng2018determination}, Ti-6246 ($\approx 4.2 \%$) \cite{ackerman2020interface}, and Ti-5-2.5-0.5 ($\approx 4 \%$).

\begin{figure}
    \centering
    \includegraphics[width= \linewidth]{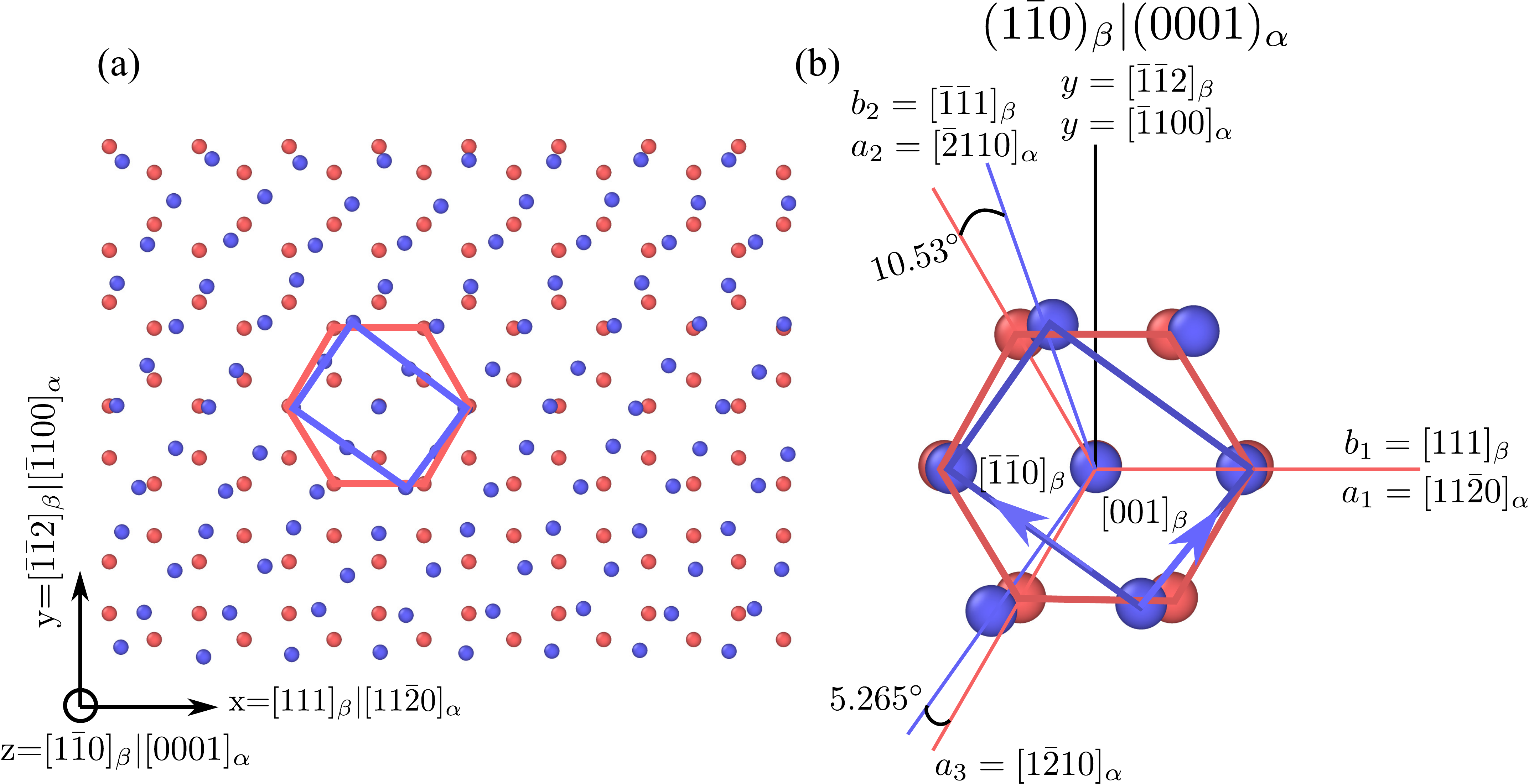}
    \caption{(a) The atomic configuration of two $(1 \bar{1} 0)_{\beta}$ planes superposed onto two $(0 0 0 1)_{\alpha}$ planes to illustrate the BOR in the dual-phase alloy studied here. Atoms on a perfect BCC lattice are shown in blue whereas those on a perfect HCP lattice are shown in red as determined by the adaptive common neighbor analysis algorithm \cite{stukowski2009visualization}. (b) The Burgers orientation variant used to describe the relationship between the $\alpha$ and the $\beta$ phases in this study.}
    \label{Fig:Burger_orientation_relationship}
\end{figure}

The $\alpha$ and $\beta$ phases are consistently observed to follow the BOR \cite{furuhara1996crystallography}. There are 12 variants of such orientation relationship from which one variant is shown in Figure \ref{Fig:Burger_orientation_relationship}, with the $[1 1 1]_{\beta}$, $[\bar{1}\bar{1} 2]_{\beta}$, and $[1 \bar{1} 0]_{\beta}$ directions in the BCC phase being parallel to the $[11\bar{2}0]_{\alpha}$, $[\bar{1} 1 0 0]_{\alpha}$, and $[0 0 0 1]_{\alpha}$, directions in the HCP phase, respectively. This orientation relationship is of importance as it leads to minimizing the interface strain and maximizing the interface coherency, thus, leading to the formation of a low energy interface between the two phases \cite{furuhara1996crystallography}.
 
%----------------------------------------------------------------------
%----------------------------------------------------------------------

To characterize the $\alpha-\beta$ interface side face dislocation structure, an $\alpha-\beta$ bi-crystal with a rectangular simulation cell was constructed with edge length of $60 a_{\alpha} \times 40 \sqrt{3} a_{\alpha} \times 140 c_{\alpha}$ (i.e., a total of 1,336,480 atoms). Periodic boundary conditions were employed along the $\textbf{x}$- and $\textbf{y}$-directions, and free surfaces boundary conditions along the $\textbf{z}$-direction, as shown in Figure \ref{Fig:misfit_interfacial_dislocations}(a). In the bi-crystal, the $\alpha$ and $\beta$ phases were oriented to follow the BOR shown in Figure \ref{Fig:Burger_orientation_relationship}.
A minimization procedure was then applied to the simulation cell using the conjugate gradient algorithm, where the components of the pressure tensor were relaxed to 0 bar by allowing the simulation cell to relax in all directions. The interfacial dislocations, including their line directions and their Burgers vectors, were then determined by the singular value decomposition of the atomic Nye tensor following \cite{dai2015automatic}. The atomic Nye tensor \AR{$\alpha_{ij}$ with \textit{i,j} = $\lbrace 1,2,3 \rbrace$ corresponding to $\lbrace x,y,z \rbrace$ respectively,} and its singular value decomposition were computed using the AADIS software \cite{yao2020aadis}. \AR{Each component of $\alpha_{ij}$ corresponds to the density of the Burgers vector projected onto the \textit{j} direction, in the plane perpendicular to the line direction along the \textit{i} direction. In this work,  $\alpha_{11}$ and $\alpha_{12}$ correspond to the screw and edge components of the Burgers vector density for a dislocation line along the \textbf{x}-direction, respectively.}

%**************************
\subsection{\AR{Critical resolved shear stress calculations } of $ 1/2 \langle 1 1 1 \rangle$ screw dislocation in the $\beta$ phase}

The critical resolved shear stress (CRSS) necessary to move the $1/2 \langle 1 1 1 \rangle$ screw dislocation on the $(1 \bar{1} 0)$ and $(1 1 \bar{2})$ planes is calculated for both the Ti$_{60}$Nb$_{40}$ alloy as well as the pure Nb case in the temperatures range from 5 to 300 K. For these calculations the simulation cell was oriented such that its edges were parallel to $\textbf{x} = [1 1 1]$, $\textbf{y} = [\bar{1} \bar{1} 2]$, and $\textbf{z} = [1 \bar{1} 0]$ for the $(1\bar{1}0)$ plane calculations and  $\textbf{x} = [1 1 1]$, $\textbf{y} = [ 1 \bar{1} 0]$, and $\textbf{z} = [1 1 \bar{2}]$ for the $(1 1 \bar{2})$ plane calculations, respectively. The simulation cell dimensions were $50 \sqrt{3} a_{\beta} \times 120 \sqrt{6} a_{\beta} \times 60 \sqrt{2} a_{\beta}$ (i.e., a total of 4,356,000 atoms) for the $(1 \bar{1} 0)$ plane orientation, and $50 \sqrt{3} a_{\beta} \times 200 \sqrt{2} a_{\beta} \times 50 \sqrt{6} a_{\beta}$ (i.e., a total of 4,340,000 atoms) for the $(1 1 \bar{2})$ plane orientation, respectively. A $1/2 \langle 1 1 1 \rangle$ screw dislocation was then introduced at the center of the simulation cell using its anisotropic elasticity displacement field with its line direction along the periodic $\textbf{x}$-direction \cite{yoo1971numerical}.

\begin{figure}
    \centering
    \includegraphics[width= \linewidth]{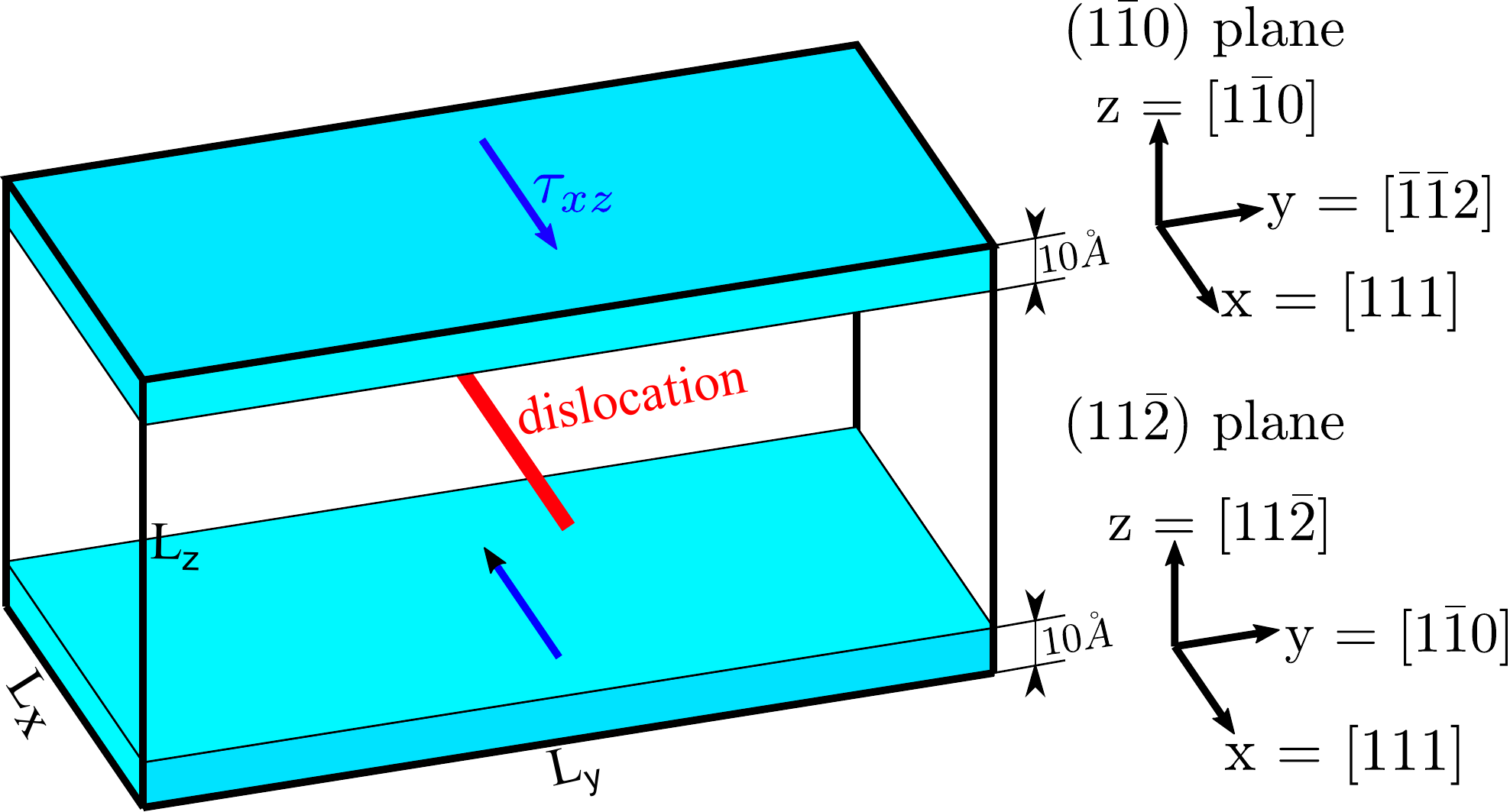}
    \caption{\AR{Schematic illustration showing the simulation setup used to compute the CRSS in the $(1\bar{1}0) $ and $(11\bar{2})$ planes in Ti$_{60}$Nb$_{40}$ alloy and pure Nb. The top and bottom 3d axes indicate the orientation of the simulation cell for the  $(1\bar{1}0) $ and $(11\bar{2})$ planes simulations, respectively. The reader is referred to the text for more detail.}}
    \label{Fig:schematic_crss}
\end{figure}

 \AR{Figure \ref{Fig:schematic_crss} shows a schematic illustration of the simulation setup used to compute the CRSS.} In both planes simulations, the dislocation line was parallel to the $\textbf{x}$-axis of the simulation cell with periodic boundary conditions employed along the $\textbf{x}$- and $\textbf{y}$-directions, while free surface boundary conditions were employed along the $\textbf{z}$-direction, respectively. Thus, mimicking the state of a periodic array of dislocations (PAD). An additional tilt of $a_{\beta} \sqrt{3}/4$ is applied in the $\textbf{y}$-direction to account for the plastic strain of the dislocation, resulting in a triclinic simulation cell \cite{osetsky2003atomic, rodney2004molecular, bacon2009dislocation}. The effects of periodicity in the dislocation glide direction on the predicted CRSS was minimized by choosing a relatively large periodic distance along the $\textbf{y}$-direction. After the introduction of the dislocation into the simulation cell a relaxation in the NVT ensemble was performed at the desired temperature for 50 ps, then another relaxation process was performed in the NPT ensemble for 50 ps to make sure that all the components of the pressure tensor on the surface of the simulation cell are at or very close to zero bar. The applied shear stress, $\tau_{xz}$, on the simulation cell was introduced by applying forces in the $\textbf{x}$-direction on atoms in a thin layer having thickness of $ \approx 10 \AA$ at both boundaries in the $\textbf{z}$-direction. The $10 \AA$ thickness is larger than twice the range of the interatomic interactions in the MEAM potential \cite{rida2022characteristics}. The forces at the top and bottom layers were equal and opposite, and the sum of the forces on either layer correspond to the applied pure shear stress. Additionally, the entire simulation cell was linearly strained by an elastic shear strain that corresponds to the applied shear stress at the onset of loading to eliminate inertial effects \cite{rao2019modeling}. These MD simulations were run for a total of 50 ps and with a time step of 1 fs. Throughout the simulations, the position of the dislocation was continuously monitored to quantify its motion as a function of time. In complex concentrated alloys the motion of the $1/2 \langle 1 1 1 \rangle$ dislocation is jerky and intermittent with some segments of the dislocation pinned at different sites due to the fluctuation in the atomic compositions in the random alloy along the dislocation line and on the glide plane \cite{varvenne2016theory, rao2019modeling,   maresca2020mechanistic, rida2022influence}. Accordingly, a lower and upper bounds of the CRSS were defined. The lower bound is the stress at which a portion of the dislocation will glide a distance of $20 \AA$ in 50 ps, whereas the upper bound is the stress at which the whole dislocation continuously glide a distance over $20 \AA$ in 50 ps.    

%----------------------------------------------------------------------
%----------------------------------------------------------------------

%**************
\subsection{Simulations of different prismatic $\langle a \rangle$ screw dislocations and their interactions with the $\alpha-\beta$ interface side face}
%The low temperature plastic behavior of high purity single crystal $\alpha$-Ti is dominated by the prismatic slip of $\langle a \rangle_{\alpha}$ dislocations \cite{clouet2015dislocation, caillard2018glide} and in $\alpha-\beta$ Ti alloys, only $\langle a \rangle_{\alpha}$ screw dislocations gliding on the prismatic planes interact with the $\alpha-\beta$ interface side face \cite{suri1999room}. Therefore, the interaction of $\langle a \rangle_{\alpha}$ prismatic screw dislocations with the $\alpha-\beta$ interface side face is expected to influence the strength and plastic behavior of two phase $\alpha-\beta$ Ti alloys. 

\begin{figure}
    \centering
    \includegraphics[width= 0.65\linewidth]{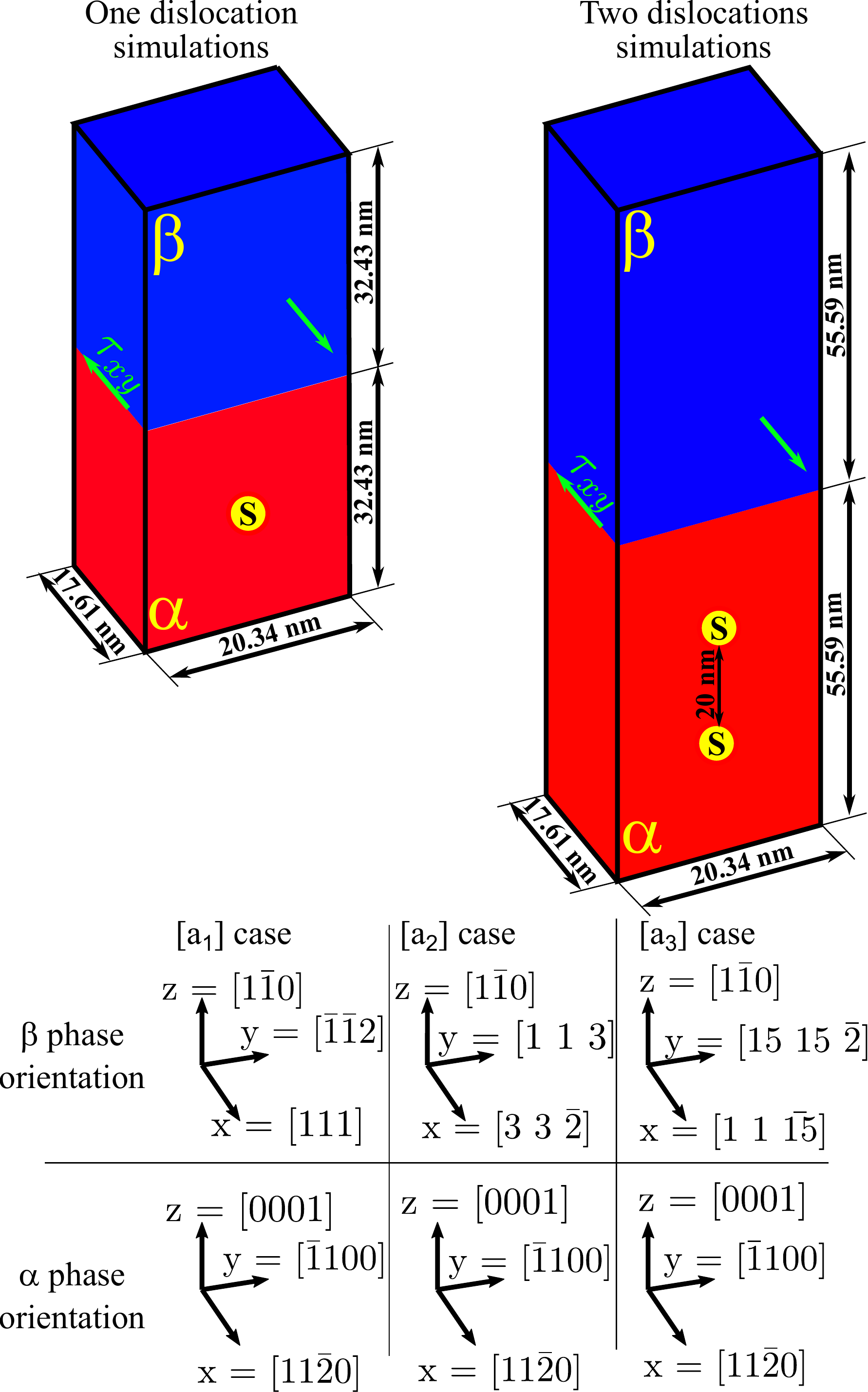}
    \caption{\AR{Schematic illustration showing the simulation setup used to study the interaction of of all three types of $\langle a \rangle_{\alpha}$ screw dislocations with the $\alpha-\beta$ interface side face. The schematic on the left describes the setup for one dislocation interacting with the interface, whereas the right schematic shows the setup for two dislocations interacting with the interface. The S in a yellow circle marks the position of the $\langle a \rangle_{\alpha}$ screw dislocation in the $\alpha$ phase. The table on the bottom of the Figure indicates the orientation of the $\alpha$ and the $\beta$ phases for each dislocation type case. The green arrows indicate the the shear stress $\tau_{xy}$ applied on the simulation cell to drive the screw dislocation toward the interface. The reader is referred to the text for more detail.}}
    \label{Fig:schematic_disl_interface}
\end{figure}

The interactions of all three type of $\langle a \rangle_{\alpha}$ screw dislocations with the $\alpha-\beta$ interface side face are studied to investigate the slip transmission mechanism and the interface strength as function of temperature between 5 and 300 K. \AR{Figure \ref{Fig:schematic_disl_interface} shows a schematic illustration of the simulation setup used to study the interaction of all three types of $\langle a \rangle_{\alpha}$ screw dislocations with the $\alpha-\beta$ interface side face.} For these simulations, first, a flat interface corresponding to the side face with the $[0 0 0 1]_{\alpha}$ normal in the $\alpha$ phase and the $[1 \bar{1} 0]_{\beta}$ normal in the $\beta$ phase was atomistically constructed in a rectangular simulation cell, as shown in Figure \ref{Fig:schematic_disl_interface}. The dimensions of the simulation cell are $60 a_{\alpha} \times 40 \sqrt{3} a_{\alpha} \times 140 c_{\alpha}$ (i.e., a total of 1,336,480 atoms) with periodic boundary conditions employed along the $\textbf{x}$-direction and free surface boundary conditions along the other two directions. In all these calculations the thickness of the $\beta$ phase is always half of the cell length in the $\textbf{z}$-direction unless otherwise stated. An $[a_1]$ screw dislocation is then inserted in the middle of the $\alpha$ phase sufficiently far from the interface (i.e., at a distance $d = 35 c_{\alpha}$ from the interface) using its anisotropic elasticity displacement field. The simulation cell was then relaxed for 50 ps using the NVT ensemble at the desired temperature, followed by another relaxation using the NPT ensemble for 50 ps to make sure that all the components of the pressure tensor on the free surfaces are at, or very close to, zero bar. The simulation cell is then loaded by a constant applied shear stress for 100 ps to drive the dislocation on the prismatic $\textbf{y}$-plane along the $\textbf{z}$-direction toward the interface. The critical stress required to transmit the $[a_1]$ screw dislocation from the $\alpha$ to the $\beta$ phase and defeat the interface is defined here as the strength of the interface. The timestep in all these simulations is 1 fs.

%----------------------------------------------------------------------
%----------------------------------------------------------------------

For the $[a_2]$ screw dislocation case, the orientation of the $\alpha$ phase of the simulation cell remains the same as discussed above. However, the orientation of the $\beta$ phase is altered such as the $\textbf{x}$-axis of the simulation cell is parallel  to the $[3 \ 3 \ \bar{2}]_{\beta}$ direction in the $\beta$ phase and the $[1 1 \bar{2} 0]_{\alpha}$ direction in the $\alpha$ phase, the $\textbf{y}$-axis is parallel to the $[1 \ 1 \ 3]_{\beta}$ in the $\beta$ phase and the $[\bar{1}100]_{\alpha}$ in the $\alpha$ phase, and the $\textbf{z}$-axis is parallel to the $[1\bar{1}0]_{\beta}$ in the $\beta$ phase and the $[0001]_{\alpha}$ in the $\alpha$ phase. This results in a $10.02^o$ misorientation between the $\textbf{x} = [a_2]$ direction in the $\alpha$ phase and the $[b_2] = [11\bar{1}]_{\beta}$ direction in the $\beta$ phase. This angle is very close to $10.53^o$ misorientation between the $[a_2]$ and $[b_2]$ Burgers vectors, as shown in Figure \ref{Fig:Burger_orientation_relationship}(b). The rest of the procedure for determining the strength of the interface to $[a_2]$ gliding screw dislocations remains similar to the procedure described for the $[a_1]$ screw dislocations above.

For the $[a_3]$ screw dislocation case, the orientation of the $\alpha$ phase remains identical to the previous two cases. However, the $\beta$ phase is oriented such that the $\textbf{x}$-axis of the simulation cell is parallel to the $[1 \ 1 \ \bar{15}]_{\beta}$ direction, the $\textbf{y}$-axis is parallel to the $[15 \ 15 \ \bar{2}]_{\beta}$ direction, and the $\textbf{z}$-axis is parallel to the $[1\bar{1}0]_{\beta}$. The result of the simulation setup remain the same. %These calculations will help understand the mechanism of transfer of slip across the side face of the interface for such dislocations, specifically, whether $\langle 001 \rangle_{\beta}$ rather than $\langle 111 \rangle_{\beta}$ dislocations are formed in the $\beta$ phase, as has been suggested from experimental observations on the broad face of the interface \cite{suri1999room, savage2004anisotropy}. 

In addition to the above calculations, a series of MD simulations were carried out to study the interaction of two $\langle a \rangle_{\alpha}$ screw dislocations on the same prismatic plane with the interface. The objective of these simulations is to focus not only on the dislocation interface interaction but also on the interaction between dislocations and residual dislocations existing on the interface due to the BOR between both phases. For these simulations, another set of MD simulations were performed for each of the three $\langle a \rangle_{\alpha}$ dislocations types, where two similar screw dislocations are initially positioned in the $\alpha$ phase on the same slip plane \AR{ as shown in Figure \ref{Fig:schematic_disl_interface}}. The orientation of the simulation cell for each dislocation type case is the same as those corresponding to a single dislocation having the same dislocation type discussed above. However, the simulation cell dimensions for these cases are $60 a_{\alpha} \times 40 \sqrt{3} a_{\alpha} \times 240 c_{\alpha}$ (i.e., a total of 2,287,680 atoms) with the $\alpha$ phase occupying the lower half of the simulation cell in the $\textbf{z}$-direction, while the $\beta$ phase the upper half, respectively. Periodic boundary conditions were employed along the $\textbf{x}$-direction and free surface boundary conditions along the other two directions. The two screw dislocations are then introduced in the $\alpha$ phase using their anisotropic displacement field, such that they are at distances of $\sim 18$ nm and $\sim 38$ nm  from the interface, respectively. These distances were chosen such that the dislocations are relatively far from the interface, far from the free surface of the $\alpha$ phase in the $z$-direction (distance to the free surface $\sim 18$ nm) to avoid any image forces, as well as far from each other.

%*********************************************
\section{Results}
\subsection{The misfit dislocations structure on the $\alpha-\beta$ interface side face}
\begin{figure}
	\centering
	\includegraphics[width= \linewidth]{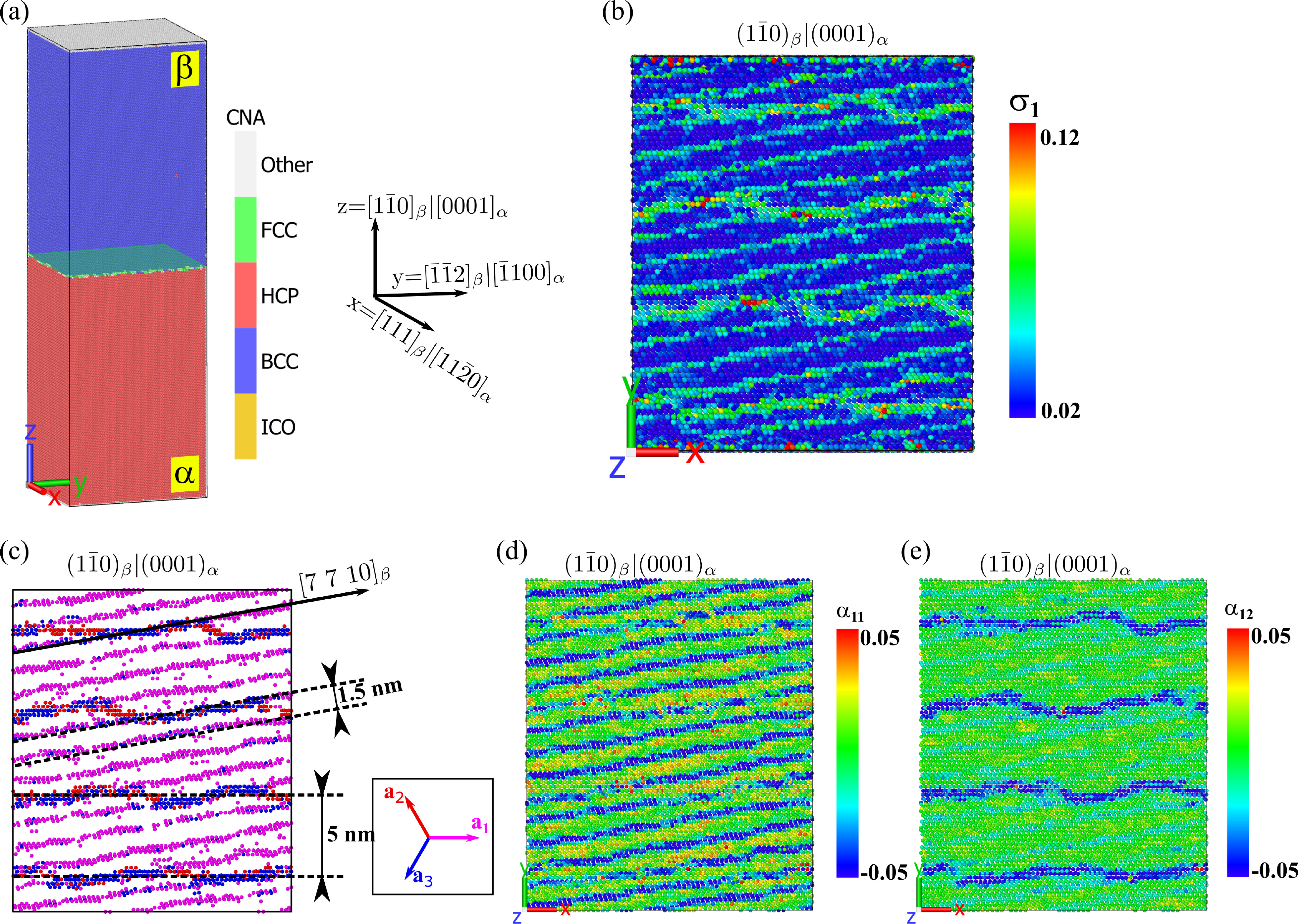}
	\caption{(a) Snapshots of the relaxed atomic simulation cell used to characterize the misfit dislocations structure on the interface side face plane highlighted by light green. The atoms are colored by their local crystal structure environment using the adaptive common neighbor analysis algorithm implemented in OVITO \cite{stukowski2009visualization}. (b) Top view of the interface plane colored by the first singular value of the atomic Nye tensor singular value decomposition \cite{dai2015automatic, yao2020aadis} where the atoms belonging to the misfit dislocation core structure have $\sigma_1>0.04$. (c) The same snapshot of the interface plane but colored with the atomic Burgers vector \cite{dai2015automatic}, the atoms belonging to the coherent region between the two phases on the interface were omitted for clarity using the above mentioned threshold of $\sigma_1$. (d) and (e) also show the same snapshot of the interface plane but colored with the $\alpha_{11}$ and $\alpha_{12}$ components of the atomic Nye tensor illustrating each set of misfit dislocations separately. The reader is referred to the text for more detail.}
	\label{Fig:misfit_interfacial_dislocations}
\end{figure}

The misfit dislocations structure on the relaxed interface were characterized using the singular value decomposition of the atomic Nye tensor \cite{dai2015automatic}. This method not only allows to determine the atomic Nye tensor but also to compute the associated atomic dislocations' Burgers vectors and line directions. Figure \ref{Fig:misfit_interfacial_dislocations}(b) shows a top view of the relaxed interface side face colored by the first singular value, $\sigma_1$, of the atomic Nye tensor. A threshold value of $\sigma_1 > 0.04$ is specified to determine the atoms belonging to a dislocation core. Two sets of misfit parallel dislocations can be clearly distinguish on the interface. In one set the dislocations are parallel with a smaller spacing, while those in the other set are more coarsely spaced. Figure \ref{Fig:misfit_interfacial_dislocations}(c) shows the misfit dislocations core structures on the interface colored by their atomic Burgers vector. The coherent atoms on the interface were omitted for clarity using the above mentioned $\sigma_1$ threshold. The Burgers vector of the finely spaced set is $[a_1] = 1/3[1 1 \bar{2} 0]_{\alpha}$ in the $\alpha$ phase, or $[b_1] = 1/2[1 1 1]_{\beta}$ in the $\beta$ phase. These dislocations are near negative screws with dislocation line direction along the $[7 \ 7  \ 10]_{\beta}$ direction, which is $\sim 10^o$ from the $[1 1 1]_{\beta}$ direction as seen in the Figure \ref{Fig:misfit_interfacial_dislocations}(c). Therefore, this direction is the invariant line direction of the $\beta$ to $\alpha$ transformation associated with the BOR. This invariant line direction matches the analytical calculations using the phenomenological theory of martensite crystallography and the topological model developed by Pond et al. \cite{pond2003comparison} (see \ref{Invariant_line_direction} for the detail of this calculation). The spacing between these dislocations is nearly $1.5$ nm. \AR{This spacing distance is bigger than the misfit dislocations core spreading on the interface as seen in Figure \ref{Fig:misfit_interfacial_dislocations}~(c). This reduces the core effects on the misfit dislocations interactions. Also, }these dislocations can be isolated on the interface using the $\alpha_{11}$ component of the atomic Nye tensor as shown in Figure \ref{Fig:misfit_interfacial_dislocations}(d). The value of $\alpha_{11}$ associated with the atoms on the misfit dislocation core is negative with an orientation of the dislocation line in agreement with the first singular value $\sigma_1$ color map confirming that these dislocation are negative near screw dislocations.

%-------------------------------------------------------------
%-------------------------------------------------------------

Figure \ref{Fig:misfit_interfacial_dislocations}(b) and (c) clearly show that the set of dislocations having coarser spacing have an average line direction along the $x = [1 1 1]_{\beta} || [1 1 \bar{2} 0]_{\alpha}$ direction. They are also kinked and have overlapping segments with the $[a_1]$ misfit screw dislocations. The Burgers vector for the kinked segments is $[a_2] = 1/3[\bar{2} 1 1 0]_{\alpha}$ or $[b_2] = 1/2[\bar{1} \bar{1} 1]_{\beta}$ as shown in Figure \ref{Fig:misfit_interfacial_dislocations}(c). However, the overlapping segments have $[a_3] = [1 \bar{2} 1 0]_{\alpha}$ or $[0 0 \bar{1}]_{\beta}$ Burgers vector, which is simply the sum of the two intersecting dislocations:
\begin{equation}
	[a_1] + [a_2] = -[a_3]
\end{equation} 

The spacing between the dislocations in this set is approximately $5$ nm (see Figure \ref{Fig:misfit_interfacial_dislocations}(c)). Additionally, the atoms belonging to the core of these misfit dislocations can be isolated using the $\alpha_{12}$ component of the atomic Nye tensor as shown in Figure \ref{Fig:misfit_interfacial_dislocations}(e). The $\alpha_{12}$ component for the atoms on the misfit dislocation core is also negative indicating that these dislocations are also negative $60^\circ$ dislocations.

\begin{figure}
	\centering
	\includegraphics[width= \linewidth]{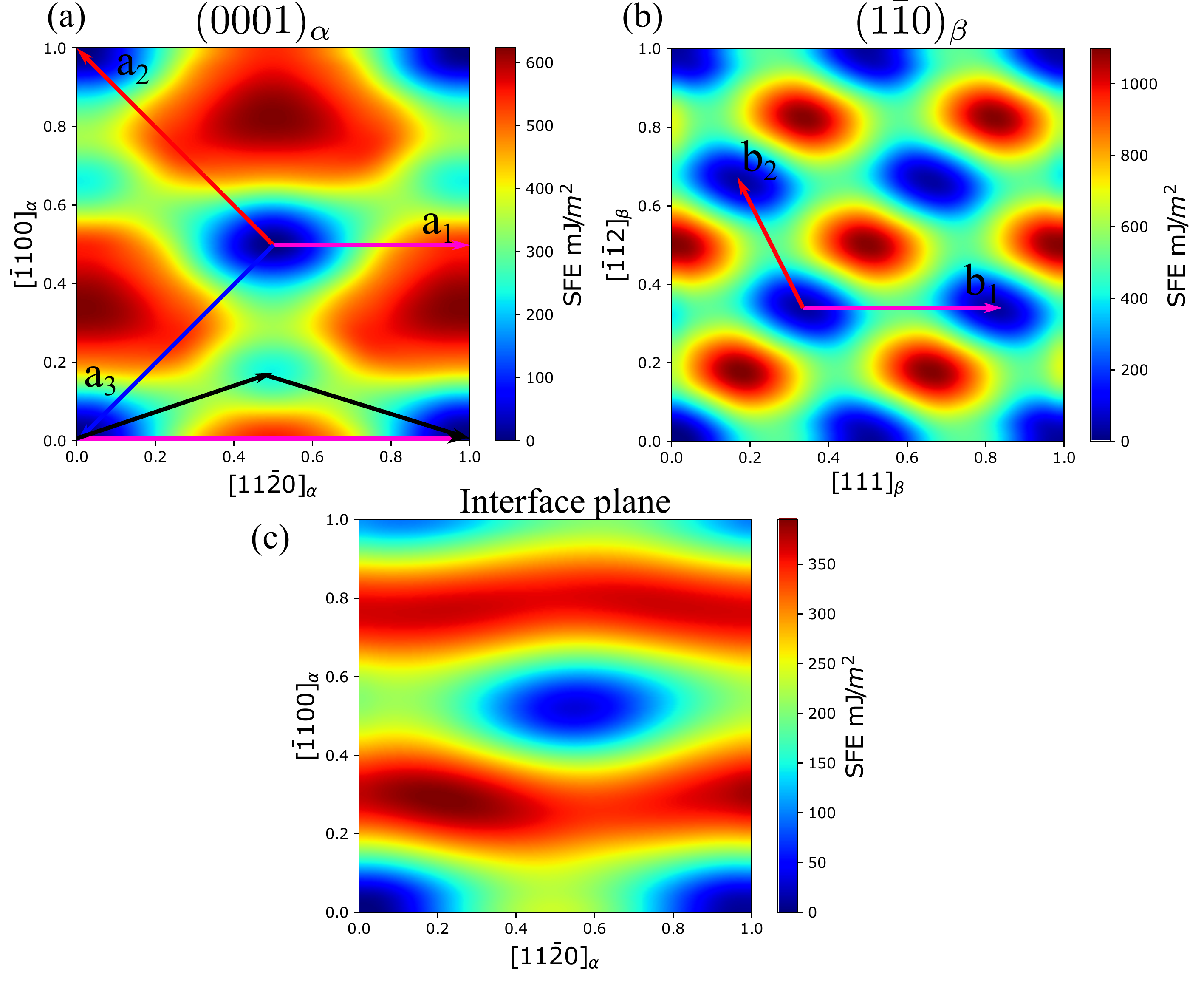}
	\caption{The GSFE surface of: (a) the $(0 0 0 1)_{\alpha}$ basal plane in pure $\alpha$-Ti; (b) the $(1 \bar{1} 0)_{\beta}$ plane in Ti$_{60}$Nb$_{40}$; and (c) the interface side face between the $\alpha$ and the $\beta$ phase as computed using the Ehemann potential. The colored arrows in (a) show the direction of the three $\langle a \rangle_{\alpha}$ directions, whereas the black arrows shows the stable SF position in which an $[a_1]$ Burgers vector will be expected to dissociate into two partials. The colored arrows in (b) indicate the direction of the two $\langle 1 1 1 \rangle$ directions.}
	\label{Fig:GSFE_surfaces}
\end{figure}

The misfit interfacial dislocations in both sets are compact on the interface plane, similar to the core structure of the $1/2\langle 111 \rangle$ screw dislocations in BCC transition metals \cite{woodward2002flexible}. To confirm the core structure of these misfit dislocations, the generalized stacking fault energy (GSFE) surface was calculated for the $(0 0 0 1)_{\alpha}$ plane in pure $\alpha$-Ti, the $(1 \bar{1} 0)_{\beta}$ plane in the Ti$_{60}$Nb$_{40}$ random alloy, and the interface plane in the $\alpha-\beta$ bi-crystal. The method for calculating the GSFE surface is similar to the method used in \cite{rida2022characteristics}. The GSFE surface for the basal $(0 0 0 1)_{\alpha}$ plane in the $\alpha$-Ti is shown in Figure \ref{Fig:GSFE_surfaces}(a), and clearly indicating a stable SF position with a SFE = 231 mJ/m$^2$. However, the GSFE surface of the $(1 \bar{1} 0)_{\beta}$ plane in the $\beta$ phase, shown in Figure \ref{Fig:GSFE_surfaces}(b), does not displace any stable SF position, confirming that $1/2\langle 1 1 1 \rangle_{\beta}$ screw dislocations will have a compact core configuration. Interestingly, the GSFE surface of the interface plane, shown in Figure \ref{Fig:GSFE_surfaces}(c),  shows a similar contour shape to the GSFE of the $(1 \bar{1} 0)_{\beta}$ plane confirming that the two sets of misfit dislocations with Burgers vector of $[b_1]$ and $[b_2]$ are compact. Finally, it is also worth mentioning that the Ehemann potential does not stabilize any dissociated basal core for the $\langle a \rangle$ screw dislocation in pure $\alpha$-Ti, in agreement with Ab-initio calculations \cite{rida2022characteristics}. This also support the fact that the interfacial misfit dislocations are compact. %Next, the core structure and critical stress of $ 1/2 \langle 111 \rangle_{\beta}$ screw dislocation in the $\beta$ phase will be determined using the Ehemann potential to assess the reliability of this potential to investigate the plastic deformation in $\alpha-\beta$ Ti alloys.

%------------------------------------------------------
%------------------------------------------------------

%*******************
\subsection{The critical resolved shear stress of $ 1/2 \langle 1 1 1 \rangle$ screw dislocations in pure Nb and Ti$_{60}$Nb$_{40}$}
\AR{In the following the focus is on computing the temperature dependence of the CRSS necessary to move a 1/2$[111]$ screw dislocation in Ti$_{60}$Nb$_{40}$ on the $(1 \bar{1} 0)$ and $(1 1 \bar{2})$ planes. The CRSS of the $(1 1 \bar{2})$ plane in the $\beta$ phase is a crucial factor that affects the resistance of the $\alpha-\beta$ interface side face to prismatic  $\langle a \rangle$ screw dislocations. In addition, the core structure of the 1/2[111]screw dislocation has an important effect on the glide of these dislocations. To this aim, before computing the CRSS, the Ehemann potential was shown to correctly reproduce the core structure of a 1/2$[111]$ screw dislocation in pure Nb and in Ti$_{60}$Nb$_{40}$ (see section S1 in supplementary materials).}

\begin{figure}
	\centering
	\includegraphics[width= 0.6\linewidth]{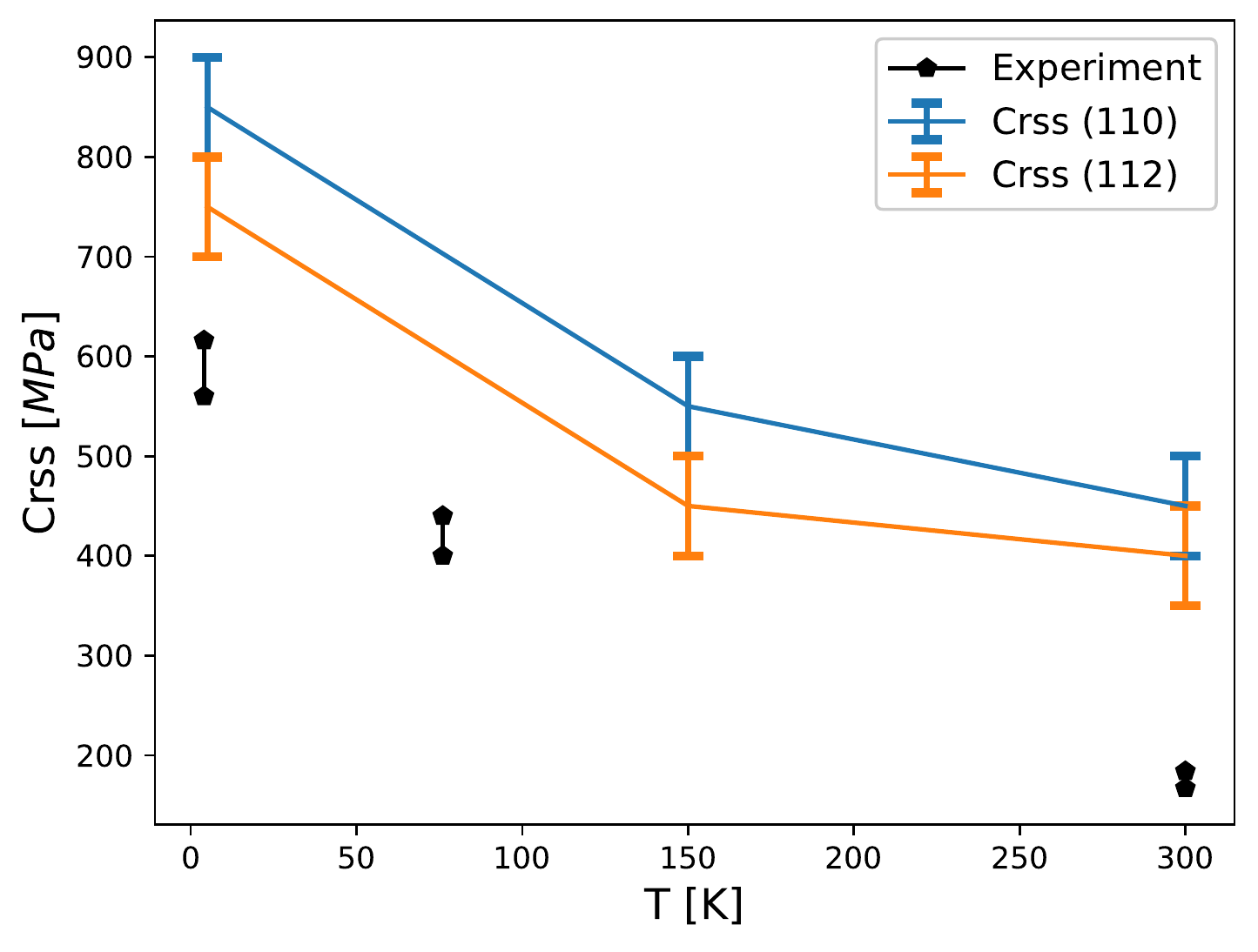}
	\caption{The CRSS for the glide of a $1/2[1 1 1]$ screw dislocation on the $(1 \bar{1} 0)$ and $(\bar{1} \bar{1} 2)$ planes in the Ti$_{60}$Nb$_{40}$ random alloy as a function of temperature. The experimentally measured CRSS at the same temperatures are also shown and were calculated as the $0.2 \%$ yield strength reported in \cite{read1978metallurgical} multiplied by a Taylor factor of $2.5$-$2.75$.}
	\label{Fig:Crss_vs_T}
\end{figure}

The upper and lower bounds of the CRSS to move a $1/2[1 1 1]$ screw dislocation on the $(1 \bar{1} 0)$ and $(1 1 \bar{2})$ planes in Ti$_{60}$Nb$_{40}$ are shown in Figure \ref{Fig:Crss_vs_T} as a function of temperature. A steep fall in the CRSS is observed between 5 and 150K for slip on both planes. This trend agrees with experimental measurements for this alloy \cite{read1978metallurgical}. However, the MD predicted CRSS values are relatively higher than experimental measurements. For temperatures above 5K, this could be attributed to strain rate effects, since the MD strain rates are several order of magnitude higher than those in the experiments. Nevertheless, at very low temperatures ($\leq 50$K) the strain rate has very low effect on the yield strength \cite{lim2015physically}. Therefore, the difference between MD and experimental strain rates cannot explain the higher CRSS values in the $\beta$ phase obtained in the current MD simulations at 5K. On the other hand, the result of the Peierls stress computed for pure Nb using the Ehemann potential is on the order of $1100 \pm 50$ MPa. This is in excellent agreement with predictions using other EAM interatomic potentials for pure Nb (\emph{cf.} \cite{zotov2021molecular}). Yet, this value is $\sim 2.5$ times higher than the experimentally predicted value of 415 MPa \cite{suzuki1995plastic}. This large difference can be explained in part by quantum effects (i.e., quantization of the crystal vibrational modes), which are not well described by interatomic potentials and was shown to be the origin of the discrepancy between simulated and experimental Peierls stresses derived from interatomic potentials or first principle calculations for other BCC metals at low temperatures \cite{proville2012quantum}. As such, the higher CRSS values in the $\beta$ phase predicted from the MD simulations at 5K can be attributed to the high Peierls stress of pure Nb predicted by the Ehemann potential.  

Figure \ref{Fig:Crss_vs_T} also shows that the CRSS for the $(1 1 \bar{2})$ planes are slightly lower than those for the $(1 \bar{1} 0)$ planes, indicating that the glide on $(1 1 2)$ planes is more favorable in this alloy. The current MD simulations also show that the screw dislocation glide is driven by the migration of kinks on different $\lbrace 1 1 0\ \rbrace$ planes, which will produce cross-kinks leading to the formation of jogs, as shown in supplementary Figure S2 \cite{rao2019modeling}. The dragging of such jogs during the motion of $1/2[1 1 1]$ screw dislocations generates prismatic debris loops (vacancies/interstitials) from the dislocation. Such debris have also been reported in MD simulations of other MPEAs \cite{rao2019modeling,maresca2020theory}. 

%This indicates that double kink nucleation does not dominate low temperature BCC MPEAs deformation, but rather the flow is driven by kink migration along the dislocation line.   

% average gliding plane
 
%------------------------------------------------------
%------------------------------------------------------

%**********************
\subsection{The strength of the $\alpha-\beta$ interface side face to gliding prismatic $\langle a \rangle_{\alpha}$ screw dislocations}

\begin{figure}
	\centering
	\includegraphics[width= \linewidth]{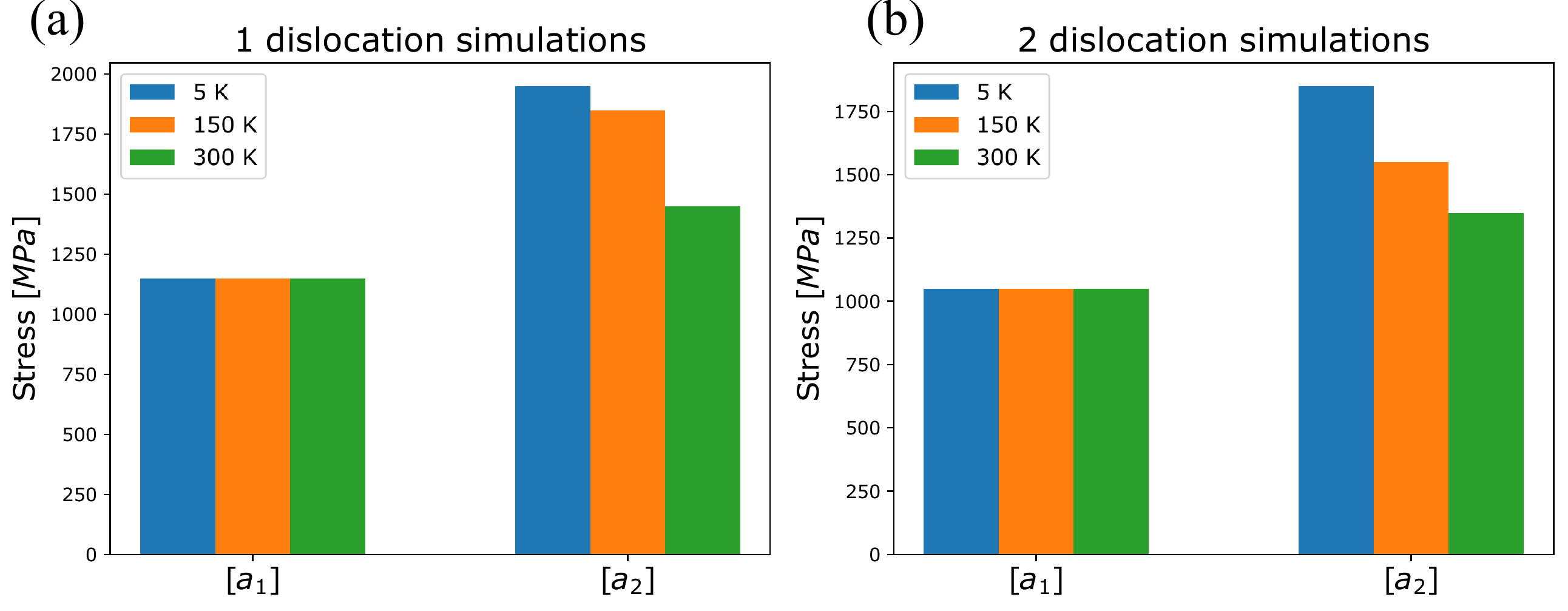}
	\caption{Strength of the $\alpha-\beta$ interface side face to $[a_1]$ and $[a_2]$ prismatic screw dislocations at different temperatures for: (a) one $\langle a \rangle_{\alpha}$ dislocation simulations; and (b) two $\langle a \rangle_{\alpha}$ dislocation pile-up simulations.}
	\label{Fig:interface_strength}
\end{figure}

Figure \ref{Fig:interface_strength}, shows the strength of the $\alpha-\beta$ interface side face to $[a_1]$ and $[a_2]$ prismatic screw dislocation slip transmission as a function of temperature for a single dislocation as well as for the case of incident of two similar dislocation. In all simulations, the strength of the interface for the transmission of $[a_1]$ prismatic screw dislocation is lower than that for the $[a_2]$ case. This indicates that the strength of the interface side face to prismatic screw $\langle a \rangle_{\alpha}$ dislocation transmission is also anisotropic, similar to the interface broad face \cite{suri1999room, savage2004anisotropy}. This anisotropy is mainly caused by the BOR and the misorientation angle between the two slip systems in the $\alpha$ and the $\beta$ phases. Additionally, these results show that the strength of the interface decrease with increasing number of dislocations in a pile-up. However, it should be noted that this stress drop is not significant and is directly related to the number of dislocations and the spacing between them in the pile-up. The repulsive interaction stress between the dislocations in the pile-up assists the first dislocation to cross the interface, and decay as $1/d$ with $d$ being the spacing between the dislocations.

While the slip transmission mechanisms for the $[a_1]$ and $[a_2]$ prismatic screw dislocations will be discussed below, it should be pointed out that in contrast to the interface broad face no transmission was observed for the $[a_3]$ dislocation in all simulations even with a dislocation pile-up. This will thus lead to higher strain hardening rate in single $\alpha-\beta$ colony oriented for prismatic $[a_3]$ slip \cite{suri1999room, savage2004anisotropy}. 

%------------------------------------------------------
%------------------------------------------------------ 

Moreover, while the strength of the interface to transmission decreases with increasing temperature for the $[a_2]$ dislocation, the strength with respect to the $[a_1]$ dislocation is surprisingly constant in the studied temperature range. When the slip planes across the interface are continuous, as in the case of the $[a_1]$ dislocation, the main factors that affect slip transmission include: (i) the stress $\sigma_{Koehler}$ induced by the difference in the dislocation line energy across the interface \cite{koehler1970attempt}; (ii) the stress $\sigma_{misfit}$ required to overcome the misfit dislocations on the interface \cite{rao2000atomistic, bacon1973effect}; (iii) the critical stress to move a $[b_1]$ dislocation on the $(\bar{1} \bar{1} 2)_{\beta}$ in the $\beta$ phase; and (iv) the stress related to the interfacial residual dislocation left on the interface, which can be neglected here since it has a very small Burgers vector $\textbf{b} = 7/600[1 1 \bar{2} 0]$ with a magnitude of $0.1 \AA$. On the other hand, $\sigma_{Koehler}$ results in an asymmetry in the slip transmission, where dislocations prefer to transmit from one phase to another if the dislocation in the latter has a lower line energy. Koehler estimated the stress required for a dislocation to overcome its elastic image stress due to the difference in the dislocation line energy across the interface as \cite{koehler1970attempt, rao2000atomistic}:
\begin{equation}
	\sigma_{Koehler} = \frac{\mu_{\alpha} (\mu_{\beta} - \mu_{\alpha})b}{4 \pi (\mu_{\beta}+\mu_{\alpha})h}
\end{equation} 
where $\mu_{\alpha}$ and $\mu_{\beta}$ are the anisotropic energy factor of an $\langle a \rangle_{\alpha}$ screw dislocation in the $\alpha$ phase, and the anisotropic energy factor of a $1/2 [1 1 1]_{\beta}$ screw dislocation in the $\beta$ phase, respectively, and $h$ is the dislocation distance from the interface. The anisotropic energy factors can be calculated from the elastic constants of the $\alpha$ and the $\beta$ phases as $\displaystyle \mu_{\alpha} = \sqrt{C_{44}(C_{11}-C_{12})/2}$ \cite{clouet2012screw} and $\displaystyle \mu_{\beta} = \sqrt{\frac{S_{11}}{(S_{11} S_{44} -S_{15}^2)S_{44}}}$, respectively \cite{hirth1966elastic}. Here, $C_{ij}$ and $S_{ij}$ are the elastic constants and compliance matrices, respectively. 

\begin{figure}
	\centering
	\includegraphics[width= \linewidth]{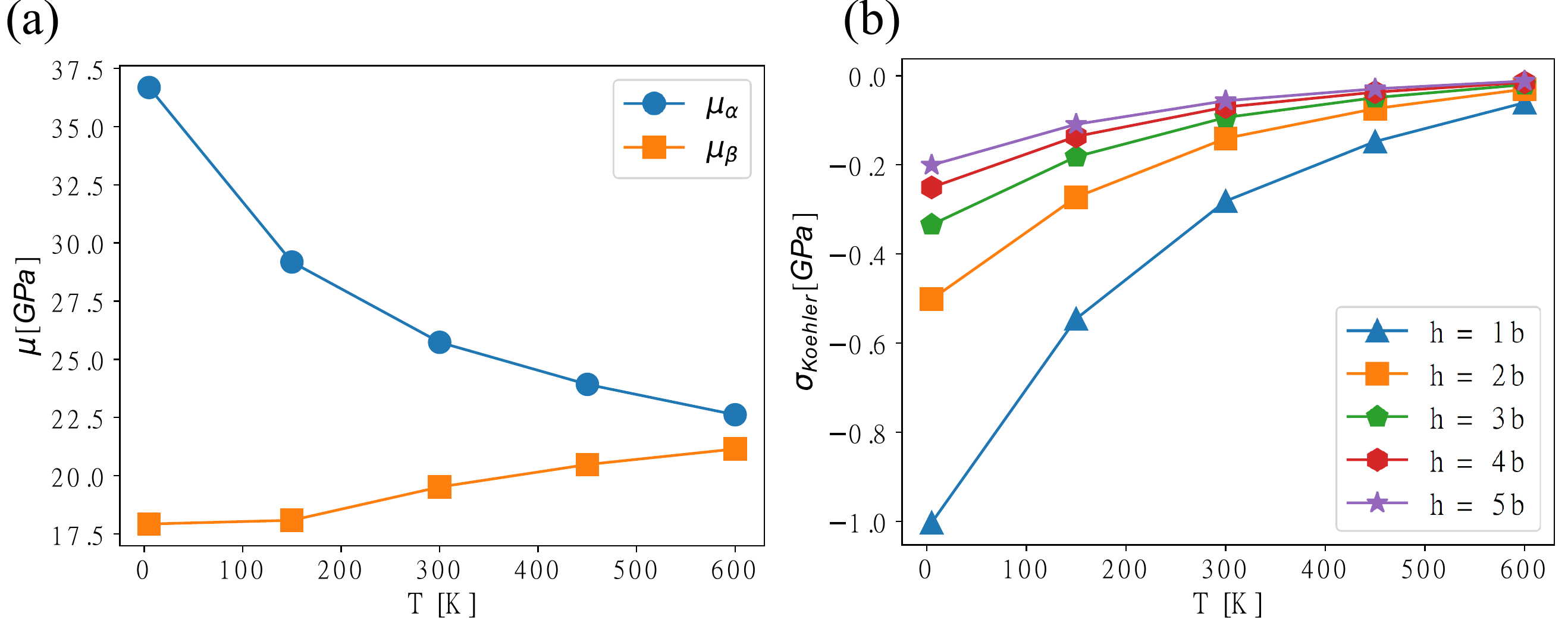}
	\caption{(a) The calculated anisotropic energy factors for an $[a_1]$ screw dislocation in the $\alpha$ phase and $[b_1]$ screw dislocation in the $\beta$ phase versus temperature as predicted using the Ehemann potential. (b) The calculated Koehler image stress of the interface on the gliding $[a_1] $ screw dislocation versus temperature for different distances, $h$, of the dislocation from the interface.}
	\label{Fig:Koehler_stres_potential}
\end{figure}

Figure \ref{Fig:Koehler_stres_potential}(a) shows the variation of the anisotropic energy factors of a screw dislocation in the $\alpha$ and the $\beta$ phases as a function of temperature. These factors were calculated from the elastic constants computed using the Ehemann potential in the temperature range from 5 to 600K (see Supplementary Figures S3 and S4). Interestingly, while the anisotropic energy factor of an $\langle a \rangle_{\alpha}$ screw dislocation decreases with increasing temperature, the energy factor of a $1/2[1 1 1]_{\beta}$ screw dislocation slightly increases. This indicates that $|\mu_{\beta} - \mu_{\alpha}|$ is not constant with temperature but rather decreasing. Hence, the absolute value of the Koehler stress affecting the glide of the dislocation will decrease with increasing temperature, as shown in Figure \ref{Fig:Koehler_stres_potential}(b) for different dislocation distances from the interface. In addition, $\mu_{\beta} \ < \ \mu_{\alpha}$ leading to a negative sign of the Koehler stress, which indicates that the dislocation is attracted to the interface when gliding on the prismatic plane in the $\alpha$ phase and repelled when gliding on the $(\bar{1} \bar{1} 2)_{\beta}$ plane in the $\beta$ phase. Therefore, a temperature increase will lead to a decrease in the interface image stress that is assisting the applied stress to drive the dislocation from the $\alpha$ to the $\beta$ phase. This Koehler stress contribution explains in part why the interface strength remains constant for $[a_1]$ dislocations in the temperature range studied here.

%------------------------------------------------------
%------------------------------------------------------ 

\begin{figure}
	\centering
	\includegraphics[width= \linewidth]{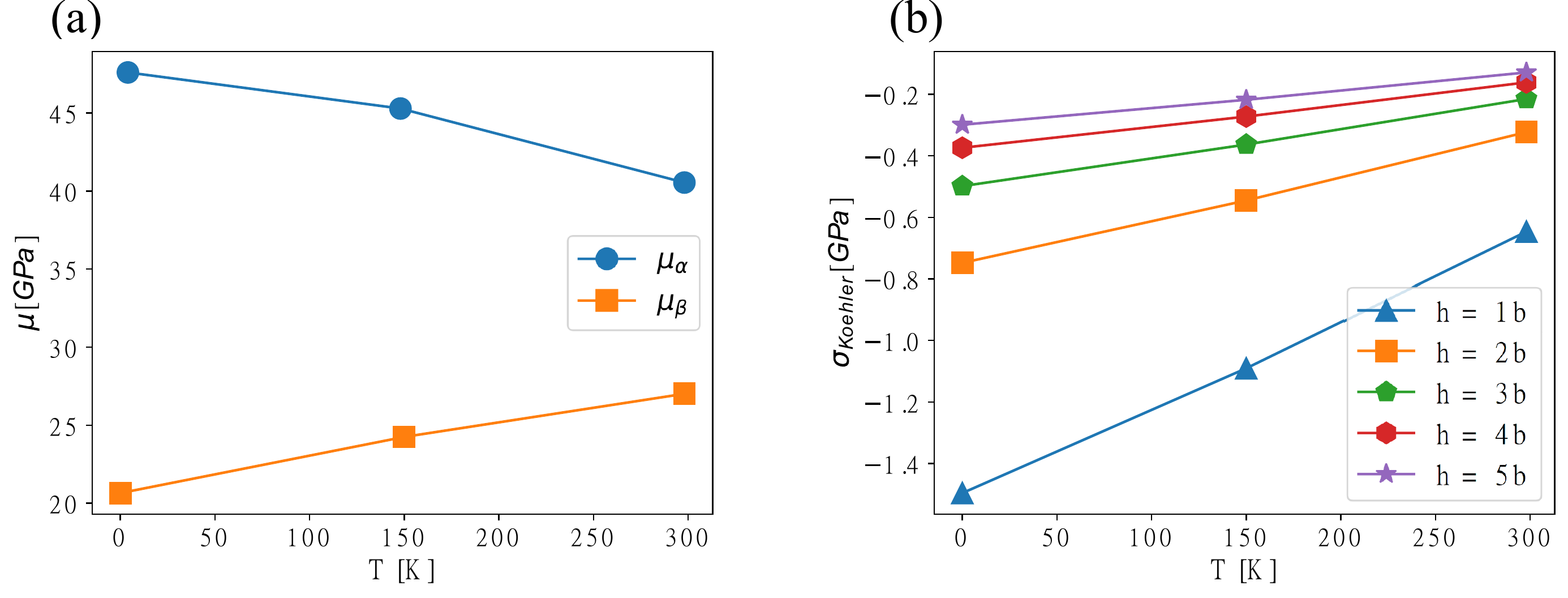}
	\caption{(a) The calculated anisotropic energy factors for an $[a_1]$ screw dislocation in the $\alpha$ phase and $[b_1]$ screw dislocation in the $\beta$ phase versus temperature using the elastic constants from the literature. (b) The calculated Koehler image stress of the interface on the gliding $[a_1] $ screw dislocation versus temperature for different distances, $h$, of the dislocation from the interface.}
	\label{Fig:Koehler_stres_exp}
\end{figure}

The Koehler stress was also calculated based on experimental data for both phases. While, the variation of the elastic constants with temperature can be found in the literature for pure $\alpha$ Ti \cite{fisher1964single}, as far as the authors are aware, only one experimental data at room temperature was reported for the Ti$_{60}$Nb$_{40}$ random alloy \cite{reid1973elastic}. Thus, the elastic constants at 0 K for the $\beta$ phase were taken from a theoretical calculations of the elastic properties of Ti$_{62.5}$Nb$_{37.5}$ performed by Fri\'ak et al \cite{friak2012theory} based on AB-initio calculations. A linear interpolation was then used to estimate the elastic constants at 150 K for the $\beta$ phase. Figure \ref{Fig:Koehler_stres_exp}(a) shows the variation of the anisotropic energy factors of the $[a_1]$ screw dislocation in the $\alpha$ and the $[b_1]$ screw dislocation in the $\beta$ phases versus temperature. The anisotropic energy factor of a screw dislocation in the $\beta$ phase slightly increases with temperature in agreement with our calculations using the Ehemann potential. The error in the anisotropic energy factors between our calculations and the data taken from the literature is between $10-35 \%$. Finally, the Koehler stress shows a similar behavior as in the interatomic potential calculations, as shown in Figure \ref{Fig:Koehler_stres_exp}(b). This further supports the argument that the temperature response observed in the current simulations is not a result of the interatomic potential used but likely of a physical origin.

%------------------------------------------------------
%------------------------------------------------------ 

\begin{figure}
	\centering
	\includegraphics[width= \linewidth]{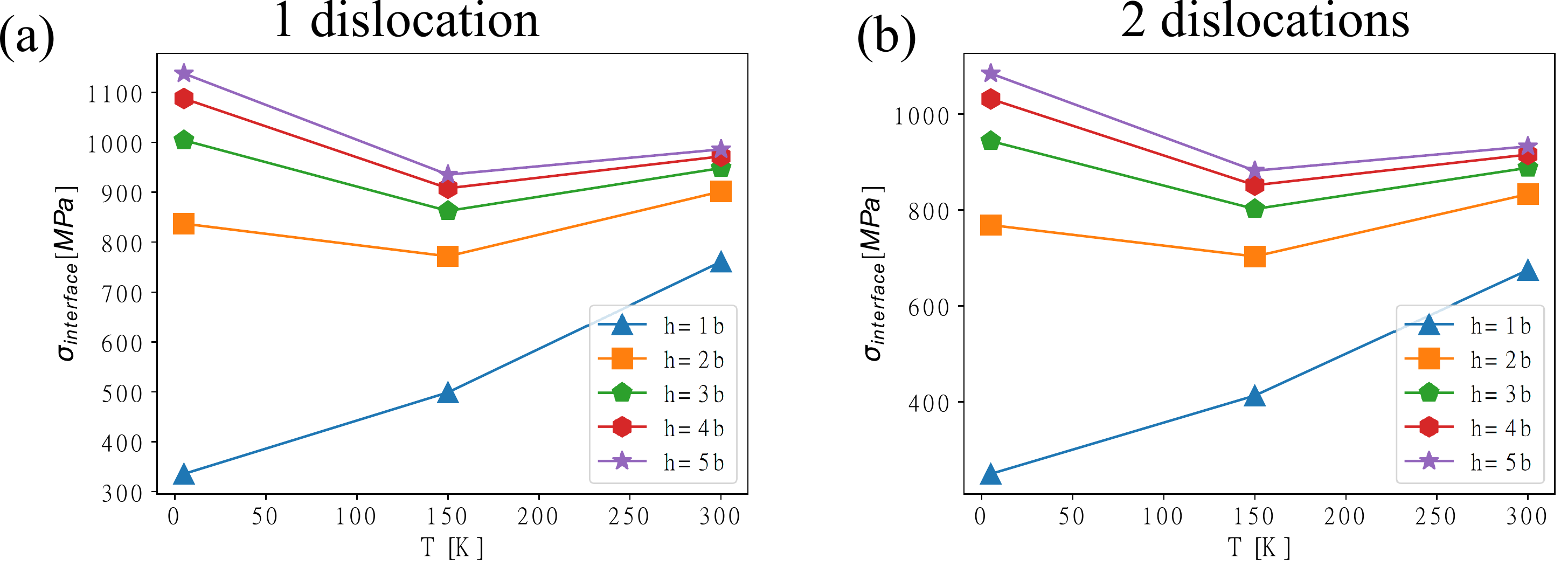}
	\caption{The calculated interface side face strength to prismatic $[a_1]$ screw dislocation versus temperature for: (a) one $[a_1]$ screw dislocation, and (b) a pile-up of two $[a_1]$ screw dislocations for different distances of the dislocation from the interface, $h$.}
	\label{Fig:analytical_interface_strength}
\end{figure}

Additionally, the contribution of the misfit dislocations to the interface strength can be approximated by the Orowan stress necessary to bow out a dislocation between rigid obstacles and is given by \cite{rao2000atomistic,bacon1973effect}:
\begin{equation}
	\sigma_{misfit} = \frac{\mu_{\beta} b}{L}
\end{equation}
where $L$ is the spacing between the misfit dislocations along the $[1 1 1]_{\beta} || [1 1 \bar{2} 0]_{\alpha}$ direction. Finally, by adding the contribution of the CRSS on the $(\bar{1} \bar{1} 2)_{\beta}$ plane shown in Figure \ref{Fig:Crss_vs_T}, the variation of the interface strength for $[a_1]$ screw dislocation can be calculated as $\sigma_{interface} = \sigma_{Koehler} + \sigma_{misfit} + CRSS_{(\bar{1} \bar{1} 2)}$, which is shown in Figure \ref{Fig:analytical_interface_strength}(a). It is observed that for a dislocation distance of $h = b$ the calculated interface strength increases with increasing temperature. For $h \geq 2b$, the variation of the interface strength with temperature is observed to be almost constant and close to the predicted MD simulation results. It should be noted that Koehler \cite{koehler1970attempt} suggested that the minimum approach distance of the dislocation to the interface should be a dislocation core width, which is for the $\langle a \rangle_{\alpha}$ prismatic screw dislocation $\approx 2b$ using the Ehemann potential \cite{rida2022characteristics}. Thus, it can be argued that the observed invariance with respect to temperature of the $\alpha-\beta$ interface strength in Ti alloys to transmission of $[a_1]$ screw dislocation could likely be physical and not an artificat of the used potential. 

%------------------------------------------------------
%------------------------------------------------------ 

In the two dislocation pileup case, the only difference from the single dislocation case is the addition of the effect of the interaction stress induced by the trailing dislocation, which acts to assist the first dislocation in transmitting across the interface. This interaction stress can be estimated as \cite{hirth1983theory}:
\begin{equation}
	\sigma_{interaction} = \frac{\mu_{\alpha} b}{2 \pi d}
\end{equation} 
where $d\approx 20$  nm is the spacing between both dislocations. By taking the variation of these parameters with temperature, the strength of the interface side face is estimated to decrease by $\approx 60-85$ MPa as shown in Figure \ref{Fig:analytical_interface_strength}(b). This estimated decrease is in good agreement with the current MD simulations, which shows a decrease of $\approx$ 100 MPa. Thus, as expected, the separation distance between dislocations in a pile-up strongly affects the slip transmission stress.  

%**************************************
\subsection{The slip transmission mechanisms}
\begin{figure}
	\centering
	\includegraphics[width= 0.8\linewidth]{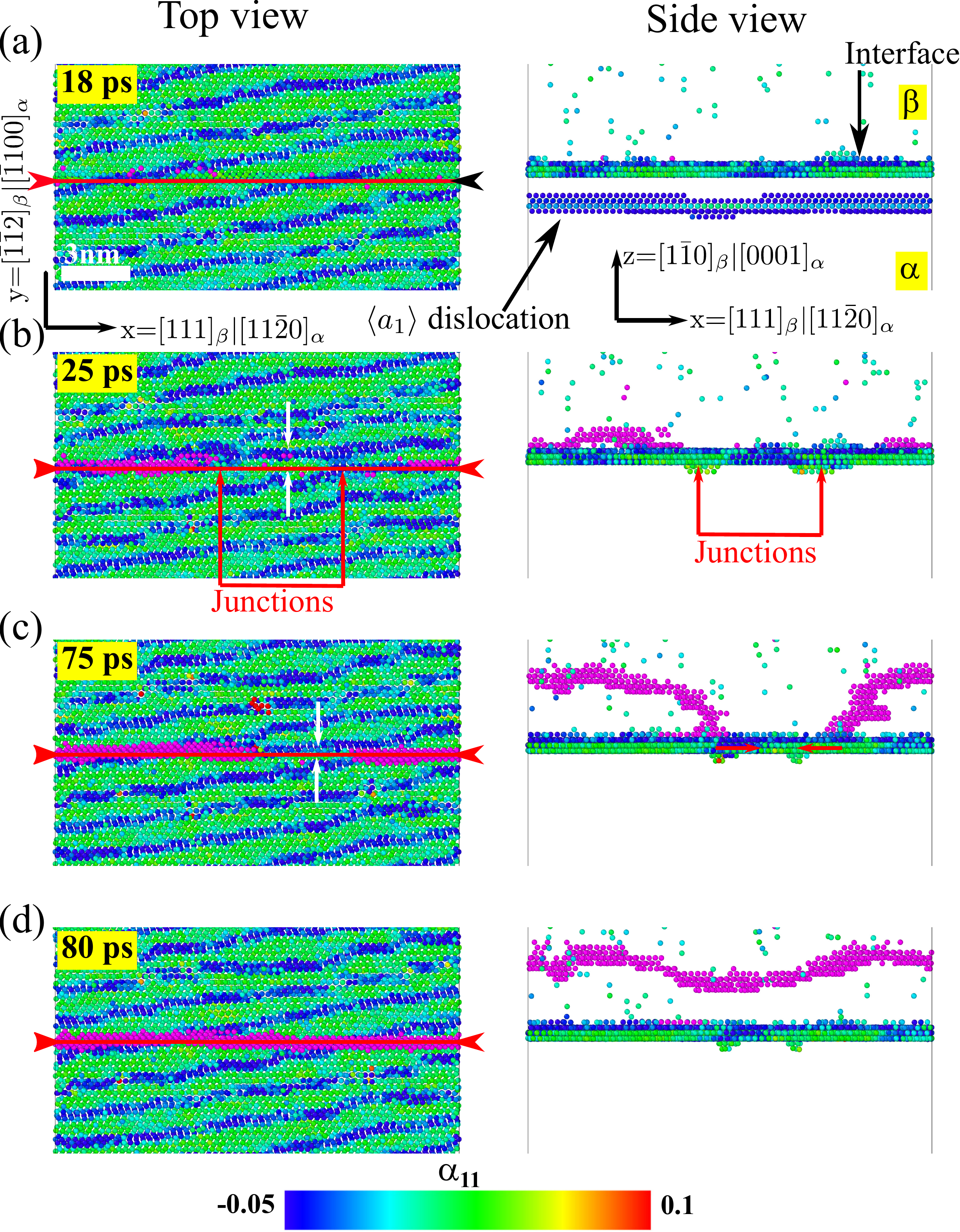}
	\caption{Sequential snapshots showing the slip transmission mechanism of the $[a_1]$ prismatic screw dislocation through the $\alpha-\beta$ interface side face at 5 K and $\sigma_{xy} = 1200$ MPa. All atoms are colored by the $\alpha_{11}$ component of the atomic Nye tensor to highlight the $[a_1]$ interfacial misfit dislocations that will interact with the incoming matrix dislocation. Only atoms on the interface plane and in the dislocation core are shown in the top and side views of the interface. Also, only atoms in the dislocation core above the exit face of the interface in the $\beta$ phase were colored in magenta to clearly follow the transmission process of the incoming matrix dislocation in both views. The position of the incoming $[a_1]$ prismatic screw dislocation in the top view of the interface is marked by a red line surrounded by two arrows (i.e, the intersection of the prismatic plane with the interface plane).}
	\label{Fig:a1_transmission_MD}
\end{figure}

To decipher the origin of the anisotropy in the strength of the interface discussed earlier, the mechanisms of slip transmission across the $\alpha-\beta$ interface for each dislocation type is analyzed. Figure \ref{Fig:a1_transmission_MD} shows sequencial atomic snapshots of the slip transmission of the $[a_1]$ prismatic screw dislocation across the $\alpha-\beta$ interface side face. The incoming negative $[a_1]$ matrix dislocation has the same sign as the interfacial misfit $[a_1]$ dislocations, as shown in Figure \ref{Fig:a1_transmission_MD}(a). When the $[a_1]$ dislocation has an opposite sign from the misfit interfacial dislocation, both dislocations will annihilate. The matrix $[a_1]$ dislocation is dissociated into two partials on the prismatic plane each having $[a_1]/2$ Burgers vectors. This dislocation glides by the formation of kink-pair mechanism on the prismatic plane as shown on the side view of Figure \ref{Fig:a1_transmission_MD}(a). As the $[a_1]$ direction is common between the interface plane and the gliding prismatic plane, the $[a_1]$ matrix dislocation will firstly cross-slip to the interface plane to interact with the $[a_1]$ misfit dislocations on the interface. Once the two partials of the $[a_1]$ dislocation are on the interface plane, dislocation junctions will form due to the repulsive interaction between the same sign interfacial and matrix dislocations, as shown in Figure \ref{Fig:a1_transmission_MD}(b). If the applied shear stress is higher than the depinning stress of the two junctions, the segments of the $[a_1]$ matrix dislocation in the coherent region between these junctions will bow out onto the $(\bar{1} \bar{1} 2)_{\beta}$ plane forming a $[b_1]$ screw dislocation in the $\beta$ phase. This happens only if these segments are aligned with the $(\bar{1} \bar{1} 2)_{\beta}$ plane. These segments are colored in magenta in Figures \ref{Fig:a1_transmission_MD}(b) and \ref{Fig:a1_transmission_MD}(c). On the other hand, a part of the dislocation between the two junctions will deviate from the $(\bar{1} \bar{1} 2)_{\beta}$ plane and rotate towards the invariant line direction to relax its core energy. This segment is marked by two white arrows on the top view of the interface plane in Figure \ref{Fig:a1_transmission_MD}(b). To unzip the complete $[b_1]$ dislocation into the $\beta$ phase, the remaining $[a_1]$ segment needs to cross slip back onto the $(\bar{1} \bar{1} 2)_{\beta}$ plane. This happens after the stress increases sufficiently on the interface plane between 25 to 75 ps, leading to a complete separation of the formed $[b_1]$ screw dislocation into the $\beta$ phase, as shown in Figure \ref{Fig:a1_transmission_MD}(c) and \ref{Fig:a1_transmission_MD}(d). The complete transmission process as observed from both views is shown in Supplementary Video V1.     

%------------------------------------------------------
%------------------------------------------------------
%------------------------------------------------------
%\begin{figure}
%	\centering
%	\includegraphics[width= \linewidth]{figures/a1_transmission_mechanism.eps}
%	\caption{}
%	\label{Fig8:schematic_a1_transmission}
%\end{figure}
%------------------------------------------------------

\begin{figure}
	\centering
	\includegraphics[width= 0.8\linewidth]{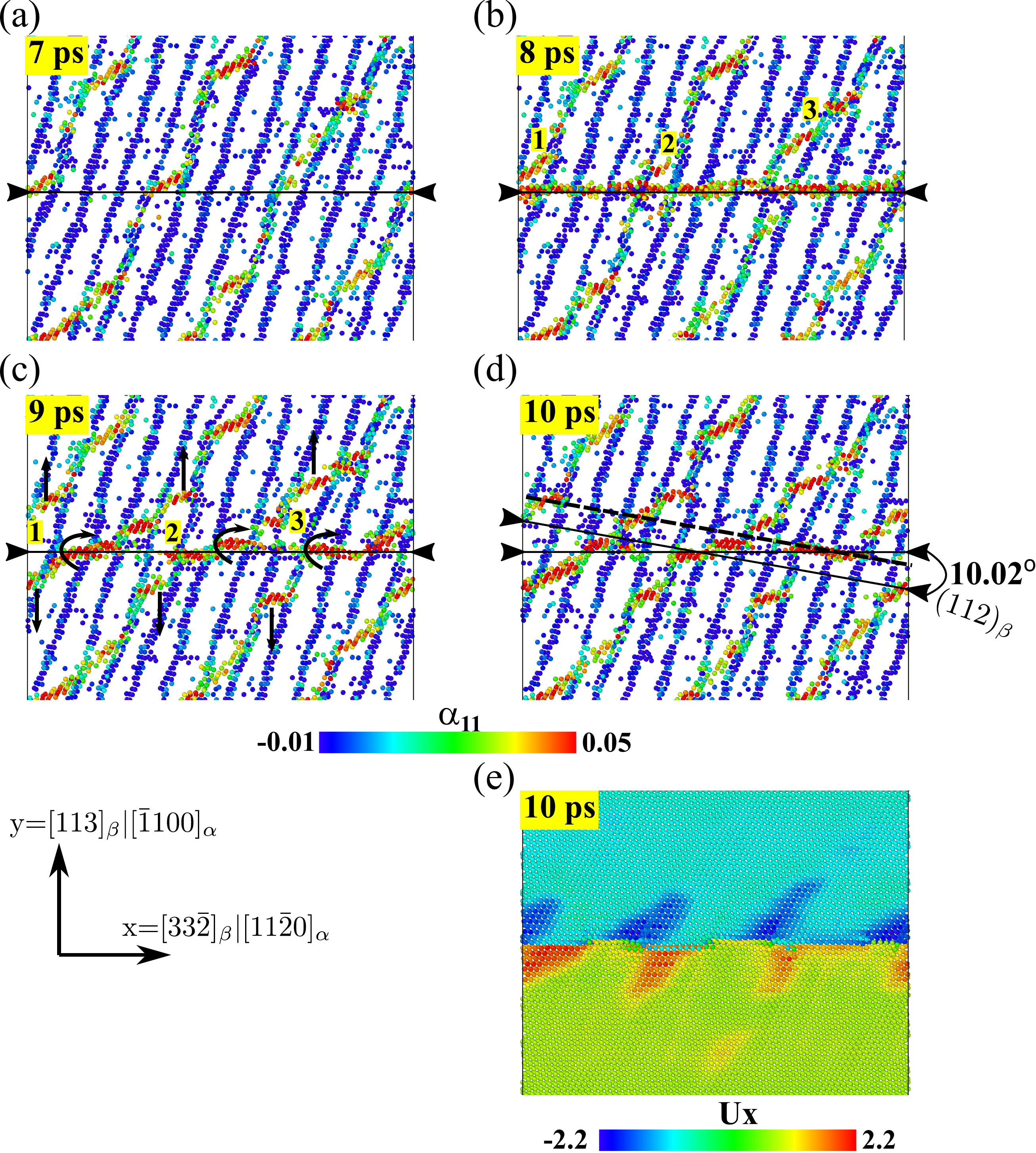}
	\caption{Sequential snapshots showing the interaction of the $[a_2]$ prismatic screw dislocation with the interface side face at 5K and $\sigma_{xy} = 1800$ MPa. The image in (a) is just before the dislocation intersects the interface; while that in (d) is after the dislocation has completely transmitted across the interface entrance face. Atoms are colored by the $\alpha_{11}$ component of the atomic Nye tensor to highlight the $[a_2]$ interfacial misfit dislocation that will interact with the incoming matrix dislocation. Only atoms on the core of the misfit dislocations are shown in the top view of the interface. The snapshot at 10 ps in (d) is also colored by the atomic displacement field computed with respect to the reference atomic configuration at 7 ps shown in (a). The atomic displacement field color map is shown in (e). The position of the incoming $[a_2]$ prismatic screw dislocation on the top view of the interface is marked by a black line surrounded by two arrows.}
	\label{Fig:a2_transmision_MD}
\end{figure}

Similar to the $[a_1]$ case, if the incoming $[a_2]$ dislocation has an opposite dislocation sign to the misfit interfacial dislocations an annihilation of this dislocation will occur at the interface. On the other hand, the slip transmission mechanism for the $[a_2]$ prismatic screw dislocation is more complex than that observed for the $[a_1]$ screw dislocation case. The interaction of an $[a_2]$ screw dislocation with the same sign $[a_2]$ misfit dislocations is shown in Figure \ref{Fig:a2_transmision_MD}. The red atoms in Figure \ref{Fig:a2_transmision_MD}(a)-(d) are the segments with positive $[a_2]$ Burgers vector on the interface as characterized by their positive $\alpha_{11}$ component of the Nye tensor. After 8 ps of applying a constant load of $\sigma_{xy} = 1800$ MPa the first partial of the positive prismatic $[a_2]$ screw dislocation just hit the entrance face of the side face. The $[a_2]$ matrix dislocation firstly cross slips to the interface plane and interacts with the $[a_2]$ segments of the misfit dislocations on the interface. The first segments to cross slip are the segments next to the $[a_2]$ misfit dislocation segments on the interface, marked by numbers 1, 2, and 3 in Figure \ref{Fig:a2_transmision_MD}(b). These segments will experience repulsive forces from the same sign $[a_2]$ interfacial dislocation segments, as represented by black arrows in Figure \ref{Fig:a2_transmision_MD}(c). This is clearly observed from the atomic displacement field color map shown in Figure \ref{Fig:a2_transmision_MD} (e), where the $[a_2]$ segments at positions 1, 2, and 3 have clearly relaxed their core energies by cross-slipping on the interface plane, while the remaining segments remain on the $\alpha$ prismatic plane. After the cross slip of the $[a_2]$ matrix dislocation segments onto the interface plane their Burgers vector will transform from $[a_2]$ to $[b_2]$, forming residual segments with the same line direction and an opposite sign to the incoming $[a_2]$ matrix dislocations. The $[b_2]$ segments will then rotate their line direction to be aligned with the $(1 1 2)_{\beta}$ plane in the $\beta$ phase with $\sim 10^\circ$ misorientation from the prismatic plane in the $\alpha$ phase as shown in Figure \ref{Fig:a2_transmision_MD}(c) and (d).

%------------------------------------------------------
%------------------------------------------------------ 

\begin{figure}
	\centering
	\includegraphics[width= \linewidth]{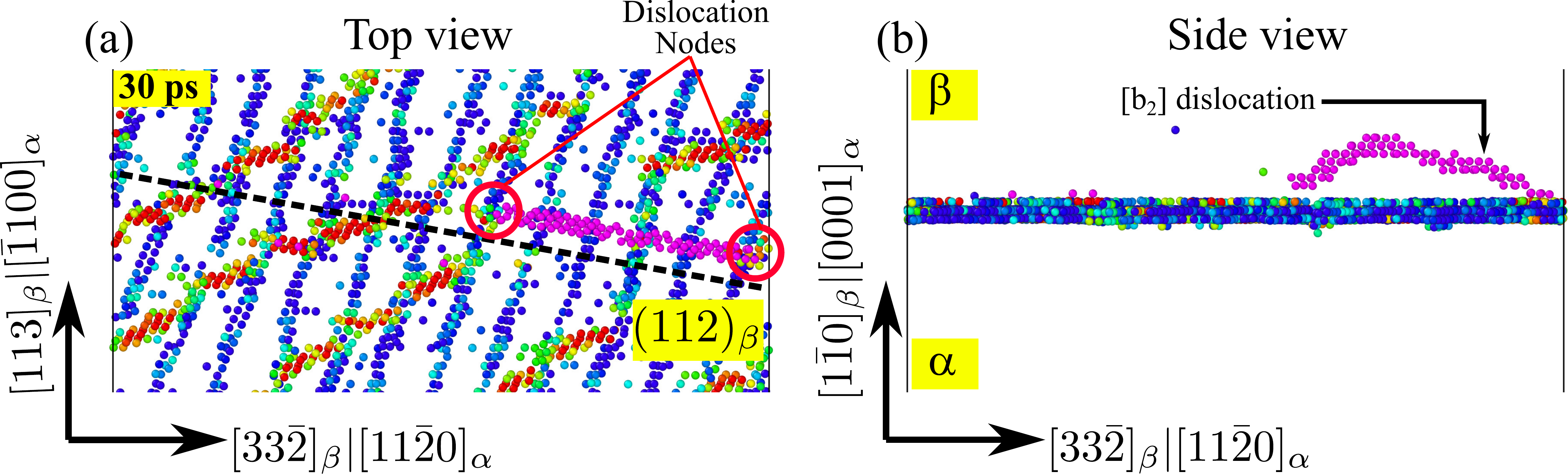}
	\caption{Top and side views of the atomic configuration of a part of the interface side face after the transmission of an $[a_2]$ prismatic screw dislocation at 5K and $\sigma_{xy} = 1800$ MPa. Only atoms in the core of the misfit and matrix dislocations are shown and colored by their $\alpha_{11}$ component of the atomic Nye tensor. Atoms above the interface plane in the $\beta$ phase are colored in magenta. The transmission of a $[b_2]$ dislocation to the $\beta$ phase only occurs after the interaction of two $[a_2]$ dislocations in 30 ps. The orientation of the $(1 1 2)_{\beta}$ plane is marked by the dashed black line in (a). }
	\label{Fig:a2_transmision_MD_continuity}
\end{figure}

The complete transmission of this $[a_2]$ dislocation only occurs after the interaction of the interface with a second $[a_2]$ dislocation, which will lead to an increase in the stress concentration on the interface. A transmission of the $[b_2]$ segments aligned with the $(1 1 2)_{\beta}$ plane on the interface to the $\beta$ phase will thus release the stress concentration on the interface. Figure \ref{Fig:a2_transmision_MD_continuity} shows the atomic configuration of the interface plane just after the transmission of the $[b_2]$ dislocation. Interestingly, the transmitted $[b_2]$ dislocation has a loop shape with screw character on the head of the loop and surrounded with two opposite edge pairs on the $(1 1 2)_{\beta}$ plane and pinned on two nodes on the interface plane. The ability of the $[b_2]$ dislocation to glide on the $(1 1 2)_{\beta}$ plane is dependent on the motion of these two nodes on the interface plane. This loop shape was also observed experimentally on the interface broad face for bowing $[b_2]$ dislocations in the $\beta$ phase gliding on the $(1 0 1)_{\beta}$ plane after the transmission of $[a_2]$ dislocations from the $\alpha$ phase to a thin $\beta$ lath \cite{savage2004anisotropy}. The complete transmission process as observed from both views is shown in the Supplementary Video V2.       

%------------------------------------------------------
%------------------------------------------------------

%------------------------------------------------------
%\begin{figure}
%	\centering
%	\includegraphics[width= \linewidth]{figures/a2_transmission_mechanism.eps}
%	\caption{}
%\end{figure}
%------------------------------------------------------

%************************
\section{Discussion}
%In this work, two sets of misfit dislocations were found on the side face interface. A finely spaced set aligned along the invariant line direction with a dislocation spacing of $\sim 1.5 nm$ and Burger's vector $[b_1]|[a_1]= 1/2[111]_{\beta}|1/3[11\bar{2}0]_{\alpha}$. The other set is coarser with a $\sim 5 nm$ spacing between two dislocations in this set. This set is also oriented on average along $[111]_{\beta}$ direction with an overlapping segments with the previous dislocation set. The Burger's vector of the overlapping segments is $-[a_3] = 1/3[\bar{1 }2 \bar{1} 0]_{\alpha} | [001]_{\beta} $, whereas the kinked segments have an $[a_2] | [b_2] = 1/3[\bar{2}110]_{\alpha} | 1/2[\bar{1}\bar{1}1]_{\beta}$ Burger's vector. 

The results  discussed above show the presence of two sets of dislocations on the $\alpha-\beta$ interface side face. This is consistent with the experimental characterization of the dislocation structure of the side face interface for $\alpha$ precipitates in Ti$-7.26 \ wt.\% $Cr \cite{ye2006dislocation}. The predicted Burgers vectors for the two dislocation sets as well as the smaller dislocation spacing between dislocation in one set are the same as those observed experimentally. However, the dislocation spacing for the coarser set is different. Ye and Zhang \cite{ye2006dislocation} found a spacing of 1.8 nm for the finely spaced set, which is consistent with our simulations. The spacing for the coarser set in their work was 9.4 nm, which is approximately twice the 5 nm spacing predicted here. Various dislocation spacings were also found experimentally on the broad face \cite{furuhara1995atomic, ye2006dislocation}. Based on these studies, the characteristics of the $\alpha - \beta$ interfacial structure, such as dislocation type and spacing, seems to depend on the chemical compositions of the two phase $\alpha-\beta$ Ti alloys \cite{zheng2018determination}. On the other hand, the same set of dislocation networks were also found in other HCP-BCC bi-crystals with BOR such as Mg-Nb \cite{shen2021mechanistic}, Zr-Nb \cite{lin2020dislocation}, and $\alpha-\beta$ Zr \cite{zhang2021structures}. \AR{In the former two studies \cite{shen2021mechanistic, lin2020dislocation}, the misfit dislocation structure was characterized using the atomically informed Frank-Bilby (AIFB) theory \cite{wang2013characterizing, wang2014interface, hirth2013interface} and the atomic Nye tensor analysis. The AIFB theory lead to three sets of partial dislocations on the interface in an ideal state, and upon relaxation the overlapped partial dislocations react and form two sets of full dislocations. Also, the atomic Nye tensor analysis confirmed that these dislocations are full $<a>$ dislocations. These misfit full dislocations are in excellent agreement with our misfit dislocations structure.}
%This difference can be explained by the difference in the lattice parameter of the $\alpha-\beta$ alloys between both works where the difference in the Burger's vector magnitude between both phases is $5\%$ in $Ti-7.26 \ wt.\% \ Cr$ vs $3.5\%$ in this work. 

The current results also show that the prismatic screw $[a_3]$ dislocations do not transfer to the $\beta$ phase in contrast to the $[a_1]$ and $[a_2]$ dislocations. This is resultant of the lack of a corresponding $1/2 \langle 1 1 1 \rangle_{\beta}$ Burgers vector for the $[a_3]$ dislocation in the $\beta$ phase according to the BOR. This will lead to increase in the strain hardening rate in single colony $\alpha - \beta$ Ti-alloys oriented for $[a_3]$ slip as shown experimentally \cite{savage2004anisotropy}. \AR{In contrast with the interface side face, a slip transmission of $[a_3]$ basal dislocations on the interface broad face was observed experimentally \cite{savage2004anisotropy}. A one to one correspondence was shown experimentally between $[a_3]$ dislocations in the $\alpha$ phase and $a_{\beta}[100]$ dislocations in the $\beta$ phase indicating a possible occurrence of slip transmission on this facet of the interface despite the fact that this dislocation reaction is not energetically favorable. On the other hand, large dislocation pile-ups is necessary for the slip transmission to occur confirming the large resistance of both sides of the interface to $[a_3]$ dislocations.}

In addition, an easy slip transmission mechanism is found in this work for the prismatic $[a_1]$ screw dislocation as seen in Figure \ref{Fig:a1_transmission_MD}. This mechanism is controlled by the depinning stress of the junctions formed by the interaction between the misfit dislocations on the interface with the incoming matrix dislocation. The easiness of this mechanism is related to the perfect alignment of the prismatic slip plane with the $(\bar{1} \bar{1} 2)_{\beta}$ plane in the BOR, and the small Burgers vector magnitude difference between $[a_1]$ and $[b_1]$ dislocations. This is also in agreement with experimental observations on single colony two phase $\alpha-\beta$ Ti-alloys oriented for $[a_1]$ slip, where the strength of the $\alpha - \beta$ interface and the strain hardening rate for $[a_1]$ dislocations were lower than that of the $[a_2]$ and $[a_3]$ dislocation types \cite{suri1999room, savage2004anisotropy}. On the other hand, the slip transmission mechanism of the $[a_2]$ dislocation is more complex and must be aided by dislocation pileup onto the side face. This complex mechanism is due to the $\sim 10^\circ$ misorientation between the two slip planes in both phases. The $[a_2]$ slip transmission mechanism is controlled by: (i) the ability of the cross slipped $[b_2]$ segments to rotate onto the $(1 1 2)_{\beta}$ slip plane in the $\beta$ phase; and (ii) the ability of the two dislocation nodes to move on the interface plane, as shown in Figure \ref{Fig:a2_transmision_MD_continuity}. These results are in agreement with the experimental work of Suri et al. \cite{suri1999room} and Savage et al. \cite{savage2004anisotropy}, where higher strength and strain hardening rates were observed for the $[a_2]$ prismatic and basal slip systems interacting with the interface broad face.    

%*************** 
\section{Conclusions}
In this work MS and MD simulations were performed to study the slip transmission of $\langle  a \rangle_{\alpha}$ prismatic screw dislocations on the $\alpha-\beta$ interface side face. The misfit dislocations structure was firstly examined, two sets of dislocations were found on the interface side face in agreement with experimental observations in Ti-Cr alloys \cite{ye2006dislocation}. One set is a finely spaced set with a Burgers vector of $[b_1] = 1/2[1 1 1]_{\beta} || [a_1] = 1/3[1 1 \bar{2} 0]_{\alpha}$ and dislocation spacing of  $\sim 1.5$ nm. The other set is a coarser spaced set with a Burgers vector of $[b_2] =1/2 [\bar{1} \bar{1} 1]_{\beta} $ in the $\beta$ phase (or $ [a_2]=1/3 [\bar{2} 1 1 0]_{\alpha}$ in the $\alpha$ phase) and dislocation spacing of $\sim 5$ nm. The intersection of dislocations from both sets leads to the formation of dislocations with a Burgers vector of $\textbf{b} = [0 0 1]_{\beta} $ in the $\beta$ phase or $ -[a_3]=[1 \bar{2} 1 0]_{\alpha}$ in the $\alpha$ phase. 

Similar to the interface broad face, a significant anisotropy of the strength of the interface side face to $[a_1]$ versus $[a_2]$ prismatic screw dislocations is found. The origin of this anisotropy is due to the relative misalignment of the slip systems between the $\alpha$ and the $\beta$ phases from the BOR. The slip transmission of the $[a_1]$ dislocation was shown to be easy and mainly controlled by the depinning stress of dislocation junctions formed due to the interaction of the misfit dislocations and the incoming $[a_1]$ matrix dislocations. Whereas, the slip transmission mechanism of the $[a_2]$ dislocation is more complex with the necessity of dislocations pile ups. This mechanism is found to be controlled by the ability of the cross slipped $[b_2]$ segments to rotate onto the $(1 1 2)_{\beta}$ slip plane on the interface and the two dislocation nodes surrounding these rotated segments to move on the interface.  

\appendix
\section{Stability of the $\beta$ phase with increasing Nb fraction}
\label{beta_phase}
\begin{figure}
	\centering
	\includegraphics[width= \linewidth]{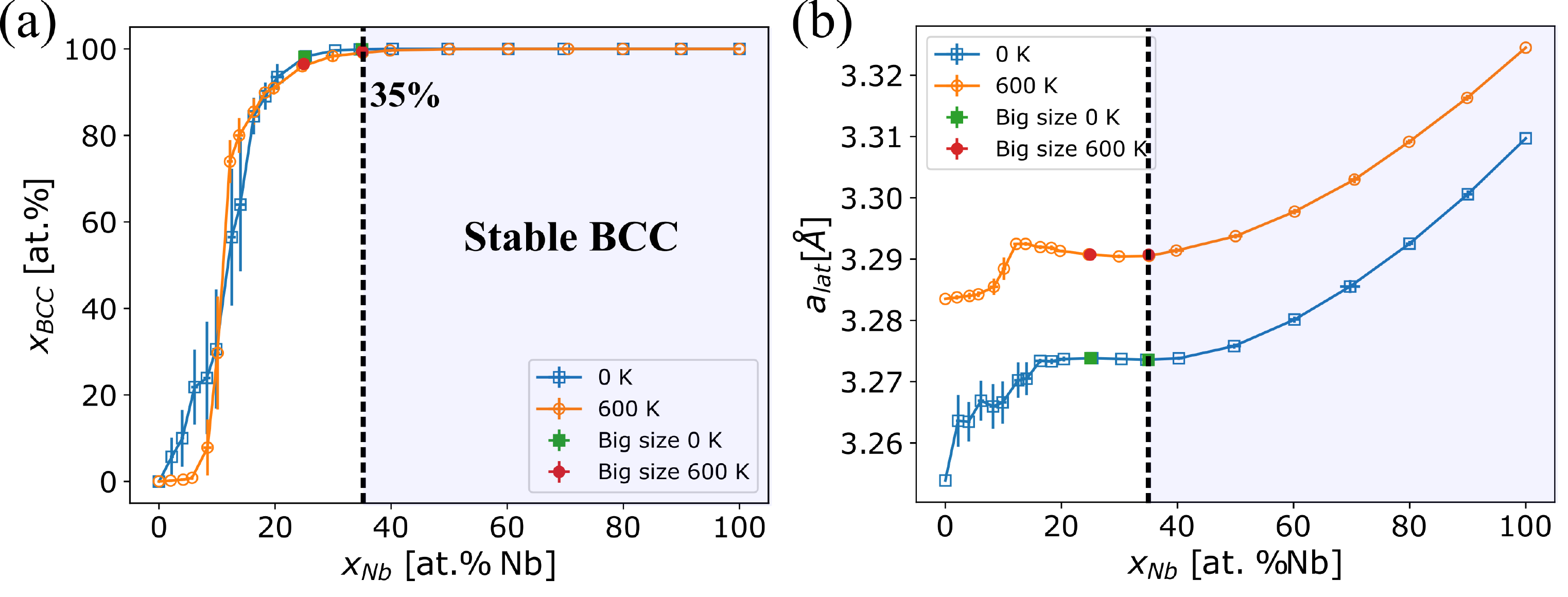}
	\caption{Variation of (a) the fraction of BCC atoms, and (b) the lattice parameter in Ti$_{100-x}$Nb$_{x}$ random alloy for different atomic fraction of Nb$_x (\%)$ at 0 and 600 K using the Ehemann potential.}
	\label{Fig:beta_phase_selection}
\end{figure}

MS and MD simulations were performed to determine the atomic fraction of Nb sufficient to stabilize the $\beta$ BCC phase in Ti-Nb random alloy. To this aim, a periodic cubic simulation cell of $10 a_{Ti} \times 10 a_{Ti} \times 10 a_{Ti}$ $\AA^3$ (i.e., 2000 atoms) was initially constructed on a BCC crystal structure in the Cartesian coordinate system with $a_{Ti}=3.25 \AA$ (i.e., the lattice parameter for $\beta$ Ti). Then a fraction of Ti atoms was randomly replaced by Nb atoms corresponding to the desired atomic percentage. Ten different random distribution simulations were performed for each fraction. After that a conjugate gradient minimization process was applied to relax all the component of the pressure tensor to zero at 0K. At 600K, the simulation cell was relaxed in NPT ensemble for 100 ps. The fraction of BCC atoms in the system was then computed using the polyhedral template matching algorithm implemented in LAMMPS \cite{larsen2016robust} and averaged over the ten realizations. Finally, the lattice parameter was also extracted and averaged over ten realizations. The effect of the simulation cell size was also investigated by repeating the same simulations as above for two atomic fraction of Nb $25 \%$ and $35 \%$ but for larger cell size of  $50 a_{Ti} \times 50 a_{Ti} \times 50 a_{Ti}$ $\AA^3$ (i.e., 250,000 atoms). No effect of the simulation cell size was found as shown in Figure \ref{Fig:beta_phase_selection}. Moreover, a fraction of at least $35 \ at. \%$ of Nb was found to be necessary to stabilize the BCC $\beta$ phase in Ti-Nb random alloy as shown in Figure \ref{Fig:beta_phase_selection}(a). Interestingly, above this fraction the variation of the lattice parameter of Ti-Nb random alloy does not follow Vegard's law as shown in Figure \ref{Fig:beta_phase_selection}(b). For this work, Ti$_{60}$Nb$_{40}$ was chosen to model the BCC $\beta$ phase in $\alpha-\beta$ Ti alloys.  

%------------------------------------------------------
%------------------------------------------------------
 
\section{Invariant line direction calculations}
\label{Invariant_line_direction}
The invariant line direction was calculated following two known methods in the literature, the phenomenological theory of martensite crystallography (PTMC), and the topological method (TM)\cite{pond2003comparison}. According to the PTMC, the $\alpha-\beta$ interface broad face or the habit plane is the invariant plane of the shape transformation $\textbf{S}$ of the $\beta$ BCC to the $\alpha$ HCP lattice at the Burgers orientation \cite{pond2003comparison, ackerman2020interface}. The habit plane is the plane containing the invariant line direction to the above transformation and the $\textbf{c} = [0 0 0 1]_{\alpha}$ direction. The variant of the BOR in Figure \ref{Fig:Burger_orientation_relationship}(b) is considered for this calculation. The deformation on the $\textbf{z}= [1 \bar{1} 0]_{\beta} || [0 0 0 1]_{\alpha}$ between both lattice is $\approx$ 0 in this direction for our considered $\alpha - \beta$ alloy transforming the problem to two dimensional case. Using this orientation, the deformation $\textbf{S}$ needed to transform the BCC lattice to HCP lattice is \cite{pond2003comparison}:

\begin{equation}
	\textbf{S} = \Omega 
	\begin{bmatrix}
		1 & \displaystyle \frac{\cos(\chi_{\beta}) - \cos(\chi_{\alpha}) }{\sin(\chi_{\beta})} & 0 \\ 0 & \displaystyle \frac{\sin(\chi_{\alpha})}{\sin(\chi_{\beta})} & 0 \\ 0 & 0 & \Omega^{-1}
	\end{bmatrix}
	=\begin{bmatrix}
		1.0355 &  -0.1831 & 0 \\ 0 & 0.9512 & 0 \\ 0 & 0 & 1
	\end{bmatrix} 
\end{equation}
where $\displaystyle \Omega = \frac{2 a_{\alpha}}{a_{\beta} \sqrt{3}}$, $\chi_{\alpha} = 60^o$ and $\chi_{\beta}=70.53^o$ are the angles between the two $\langle a \rangle$ and two $\langle b \rangle_{\beta}$ Burgers vectors in the $\alpha$ and $\beta$ phases, respectively. This deformation tensor can be decomposed to a rigid body rotation $\textbf{R}_1$ and a lattice transformation $\textbf{B}$ where:
\begin{equation}
	\textbf{R}_1 = \begin{bmatrix}
		0.9958 &  -0.0918 & 0 \\ 0.0918 & 0.9958 & 0 \\ 0 & 0 & 1
	\end{bmatrix} 
\end{equation}
and 
\begin{equation}
	\textbf{B} = \begin{bmatrix}
		1.0311 &  -0.095 & 0 \\ -0.095 & 0.9639 & 0 \\ 0 & 0 & 1
	\end{bmatrix} 
\end{equation} 
where $\textbf{R}_1$ is a rotation around the $\textbf{z}$-axis by $5.265^o$. The principal strains and directions of $\textbf{B}$ can be determined by diagonalising this matrix:
\begin{equation}
	\bm{B_d} = \begin{bmatrix}
		1.0983 &  0 & 0 \\ 0 & 0.8967 & 0 \\ 0 & 0 & 1
	\end{bmatrix} 
\end{equation}   
where the principal strains are $0 \%$ along the $\textbf{z} = [1 \bar{1} 0]_{\beta}$ direction, $9.83 \%$ along the $[1 1 0]_{\beta}$ direction and $-10.33 \%$ along the $[0 0 1]_{\beta}$ direction. The orientation of the invariant line $\textbf{v}$ can thus be found by solving \cite{ackerman2020interface}:
\begin{eqnarray}
	\textbf{(Sv)}^T \textbf{Sv} = \textbf{v}^T \textbf{B}^T \textbf{B v}=\textbf{v}^T \textbf{v}
\end{eqnarray}
where the superscript $T$ indicates the transpose of the matrix. As $\textbf{v}$ is in the (x,y) plane, the z component of $\textbf{v}$ is equal to 0. The above equation can thus be solved numerically to find the invariant line $\textbf{v} = [7 \ 7 \ 10.153]_{\beta}$ with an angle of $\sim 10.5^o $ from the $[1 1 1]_{\beta}$ direction, which matches perfectly the invariant line direction found in our simulations. The rotation between $\textbf{v}$ (the invariant vector) and $\textbf{Bv}$ (rotated but undistorted vector) is denoted $\textbf{R}_2$. $\textbf{R}_2$ is a rotation around the $\textbf{z}$-axis by $5.782^o$, calculated from the dot product between $\textbf{v}$ and $\textbf{Bv}$ by taking the relative coordinates of $\textbf{v} =(0.9834, 0.1815,0)$ in the above mentioned $\beta$ coordinate system. This will give us the final angle of deviation from the BOR, which is the difference between the rotation $\textbf{R}_2$ and $\textbf{R}_1$  as $0.517^o$. This rotation is close to that found experimentally for Ti-5553 \cite{zheng2018determination} and Ti-6246 \cite{ackerman2020interface}.

In the TM formulation, the habit plane consists of coherent prismatic terraces reticulated by arrays of interfacial defects (i.e. disconnections) ensuring that the interface exhibit no long range displacement field \cite{pond2003comparison}. In this theory a reference state is defined where the misfit strain $\varepsilon_{xx}$ between both phases is shared equally to match along the $[1 1 1]_{\beta} || [1 1 \bar{2} 0]_{\alpha}$. The Burgers vector of the disconnections $\textbf{b}$ is given by the difference between $\textbf{t}_{\alpha} = 1/3[\bar{1} 2 \bar{1} 0]_{\alpha}$ and $\textbf{t}_{\beta} = [0 0 1]_{\beta}$. The $y$  component of $\textbf{b}$  is $b_y = h_{\beta} - h_{\alpha}$, where $h_{\beta}$, and $h_{\alpha}$  the step heights of the $\beta$ and the $\alpha$ phases, respectively. The $x$  component of $\textbf{b}$ is given by $b_x = a_{\beta}/\sqrt{3}(1+\varepsilon_{xx}/2) - a_{\alpha}/2(1-\varepsilon_{xx}/2)$. The step heights of the $\alpha$ and $\beta$ phases can be calculated using lattice parameters as:
\begin{eqnarray}
    h_{\alpha} &=& a_{\alpha} sin(\chi_{\alpha})  \nonumber \\ 
    h_{\beta} &=& \frac{\sqrt{3}}{2} a_{\beta} sin(\chi_{\beta}) 
\end{eqnarray}

Following Pond et al \cite{pond2003comparison}, the inclination of the habit plane containing the invariant line direction from the $[1 1 1]_{\beta}$ direction is ($\omega = \theta - \phi$), where $\theta$ is the ledge inclination angle and $\phi$ the tilt angle from the Burgers orientation. $\theta$ and $\phi$ can be calculated by solving: 
\begin{eqnarray}
    \varepsilon_{xx} &=& \frac{b_x \tan(\theta) + b_y \tan^2 (\theta)}{h_{\alpha}} \\
    \phi &=& 2 \sin^{-1} \left(  \frac{ (\cos(\theta) - b_x \sin(\theta)) \sin(\theta)}{2 h_{\alpha}} + \frac{\varepsilon_{xx} \sin(\theta) \cos(\theta)}{2}\right)
\end{eqnarray}
where, $ \displaystyle \varepsilon_{xx} = \frac{a_{\alpha} - a_{\beta} \sqrt{3}/2}{\langle a \rangle}$, and $ \displaystyle \langle a \rangle =  \frac{a_{\alpha} + a_{\beta} \sqrt{3}/2}{2}$. Finally, the calculated orientation of the invariant line direction or the habit plane with respect to $[1 1 1]_{\beta}$ is $ \approx 10^o$  and the deviation angle from the Burgers orientation relationship is $0.518^o$ in agreement with the PTMC.

%\section{Stacking fault energy calculations for the $Ti_{60}Nb_{40}$}

\section*{Acknowledgements}
This research was sponsored by the Multiscale Structural Mechanics and Prognosis program at the Air Force Office of Scientific Research (project number: FA9550-21-1-0028). Some simulations were conducted at the Advanced Research Computing at Hopkins (ARCH) core facility (rockfish.jhu.edu), which is supported by an NSF grant number OAC-1920103. Some simulations were also conducted using the Extreme Science and Engineering Discovery Environment (XSEDE) Expanse supercomputer at the San Diego Supercomputer Center (SDSC) through allocations TG-MAT210003, and TG-MAT220013. XSEDE is supported by National Science Foundation grant number ACI-1548562.

\section*{References}
\bibliography{biblio_paper} % name your BibTeX data base

\begin{thebibliography}{10}
\expandafter\ifx\csname url\endcsname\relax
  \def\url#1{\texttt{#1}}\fi
\expandafter\ifx\csname urlprefix\endcsname\relax\def\urlprefix{URL }\fi
\expandafter\ifx\csname href\endcsname\relax
  \def\href#1#2{#2} \def\path#1{#1}\fi

\bibitem{lutjering2000microstructure}
G.~L{\"u}tjering, J.~Williams, A.~Gysler, Microstructure and mechanical properties of titanium alloys, in: Microstructure And Properties Of Materials: (Volume 2), 2000, pp. 1--77.

\bibitem{furuhara1996crystallography}
T.~Furuhara, S.~Takagi, H.~Watanabe, T.~Maki, Crystallography of grain boundary $\alpha$ precipitates in a $\beta$ titanium alloy, Metallurgical and Materials Transactions A 27~(6) (1996) 1635--1646.

\bibitem{ye2004tem}
F.~Ye, W.-Z. Zhang, D.~Qiu, A {TEM} study of the habit plane structure of intragrainular proeutectoid $\alpha$ precipitates in a {Ti}--7.26 wt\% {Cr} alloy, Acta Materialia 52~(8) (2004) 2449--2460.

\bibitem{ye2006dislocation}
F.~Ye, W.-Z. Zhang, Dislocation structure of non-habit plane of $\alpha$ precipitates in a {Ti}--7.26 wt.\% {Cr} alloy, Acta Materialia 54~(4) (2006) 871--879.

\bibitem{zheng2018determination}
Y.~Zheng, R.~E. Williams, G.~B. Viswanathan, W.~A. Clark, H.~L. Fraser, Determination of the structure of $\alpha$-$\beta$ interfaces in metastable $\beta$-{Ti} alloys, Acta Materialia 150 (2018) 25--39.

\bibitem{pond2003comparison}
R.~Pond, S.~Celotto, J.~Hirth, A comparison of the phenomenological theory of martensitic transformations with a model based on interfacial defects, Acta Materialia 51~(18) (2003) 5385--5398.

\bibitem{zhang2021study}
Y.-S. Zhang, J.-Y. Zhang, W.-Z. Zhang, A study of crystallography of $\alpha$ precipitates in a {Ti}-8 wt\% {Fe} alloy, Materials Characterization 178 (2021) 111193.

\bibitem{menon1986interfacial}
E.~S.~K. Menon, H.~Aaronson, Interfacial structure of widmanstatten plates in a {Ti-Cr} alloy, Acta Metallurgica 34~(10) (1986) 1975--1981.

\bibitem{suri1999room}
S.~Suri, G.~Viswanathan, T.~Neeraj, D.-H. Hou, M.~Mills, Room temperature deformation and mechanisms of slip transmission in oriented single-colony crystals of an $\alpha$/$\beta$ titanium alloy, Acta Materialia 47~(3) (1999) 1019--1034.

\bibitem{savage2004anisotropy}
M.~Savage, J.~Tatalovich, M.~Mills, Anisotropy in the room-temperature deformation of $\alpha$--$\beta$ colonies in titanium alloys: role of the $\alpha$--$\beta$ interface, Philosophical Magazine 84~(11) (2004) 1127--1154.

\bibitem{zhao2019slip}
P.~Zhao, C.~Shen, M.~F. Savage, J.~Li, S.~R. Niezgoda, M.~J. Mills, Y.~Wang, Slip transmission assisted by shockley partials across $\alpha$/$\beta$ interfaces in {Ti}-alloys, Acta Materialia 171 (2019) 291--305.

\bibitem{thompson2022lammps}
A.~P. Thompson, H.~M. Aktulga, R.~Berger, D.~S. Bolintineanu, W.~M. Brown, P.~S. Crozier, P.~J. in't Veld, A.~Kohlmeyer, S.~G. Moore, T.~D. Nguyen, et~al., Lammps-a flexible simulation tool for particle-based materials modeling at the atomic, meso, and continuum scales, Computer Physics Communications 271 (2022) 108171.

\bibitem{ehemann2017force}
R.~C. Ehemann, J.~W. Wilkins, Force-matched empirical potential for martensitic transitions and plastic deformation in {Ti}-{Nb} alloys, Physical Review B 96~(18) (2017) 184105.

\bibitem{rida2022characteristics}
A.~Rida, S.~I. Rao, J.~A. El-Awady, Characteristics of< a> screw dislocations and their slip on prismatic and pyramidal planes in pure $\alpha$ titanium from atomistic simulations, Materialia (2022) 101503.

\bibitem{clouet2015dislocation}
E.~Clouet, D.~Caillard, N.~Chaari, F.~Onimus, D.~Rodney, Dislocation locking versus easy glide in titanium and zirconium, Nature materials 14~(9) (2015) 931--936.

\bibitem{moffat1988stable}
D.~Moffat, U.~Kattner, The stable and metastable {Ti}-{Nb} phase diagrams, Metallurgical Transactions A 19~(10) (1988) 2389--2397.

\bibitem{ackerman2020interface}
A.~K. Ackerman, V.~A. Vorontsov, I.~Bantounas, Y.~Zheng, Y.~Chang, T.~McAuliffe, W.~A. Clark, H.~L. Fraser, B.~Gault, D.~Rugg, et~al., Interface characteristics in an $\alpha$+ $\beta$ titanium alloy, Physical Review Materials 4~(1) (2020) 013602.

\bibitem{stukowski2009visualization}
A.~Stukowski, Visualization and analysis of atomistic simulation data with ovito--the open visualization tool, Modelling and Simulation in Materials Science and Engineering 18~(1) (2009) 015012.

\bibitem{dai2015automatic}
F.-Z. Dai, W.-Z. Zhang, An automatic and simple method for specifying dislocation features in atomistic simulations, Computer Physics Communications 188 (2015) 103--109.

\bibitem{yao2020aadis}
B.~Yao, R.~Zhang, Aadis: An atomistic analyzer for dislocation character and distribution, Computer Physics Communications 247 (2020) 106857.

\bibitem{yoo1971numerical}
M.~Yoo, B.~Loh, Numerical calculation of elastic properties for straight dislocations in anisotropic crystals., Tech. rep., Oak Ridge National Lab., Tenn. (1971).

\bibitem{osetsky2003atomic}
Y.~N. Osetsky, D.~J. Bacon, An atomic-level model for studying the dynamics of edge dislocations in metals, Modelling and Simulation in Materials Science and Engineering 11~(4) (2003) 427.

\bibitem{rodney2004molecular}
D.~Rodney, Molecular dynamics simulation of screw dislocations interacting with interstitial frank loops in a model fcc crystal, Acta Materialia 52~(3) (2004) 607--614.

\bibitem{bacon2009dislocation}
D.~Bacon, Y.~Osetsky, D.~Rodney, Dislocation--obstacle interactions at the atomic level, Dislocations in Solids 15 (2009) 1--90.

\bibitem{rao2019modeling}
S.~Rao, B.~Akdim, E.~Antillon, C.~Woodward, T.~Parthasarathy, O.~Senkov, Modeling solution hardening in {BCC} refractory complex concentrated alloys: {NbTiZr}, {Nb1. 5TiZr0. 5} and {Nb0. 5TiZr1. 5}, Acta Materialia 168 (2019) 222--236.

\bibitem{varvenne2016theory}
C.~Varvenne, A.~Luque, W.~A. Curtin, Theory of strengthening in {FCC} high entropy alloys, Acta Materialia 118 (2016) 164--176.

\bibitem{maresca2020mechanistic}
F.~Maresca, W.~A. Curtin, Mechanistic origin of high strength in refractory {BCC} high entropy alloys up to 1900k, Acta Materialia 182 (2020) 235--249.

\bibitem{rida2022influence}
A.~Rida, E.~Martinez, D.~Rodney, P.-A. Geslin, Influence of stress correlations on dislocation glide in random alloys, Physical Review Materials 6~(3) (2022) 033605.

\bibitem{woodward2002flexible}
C.~Woodward, S.~Rao, Flexible ab initio boundary conditions: Simulating isolated dislocations in {BCC} {Mo} and {Ta}, Physical Review Letters 88~(21) (2002) 216402.

\bibitem{read1978metallurgical}
D.~Read, Metallurgical effects in niobium-titanium alloys, Cryogenics 18~(10) (1978) 579--584.

\bibitem{lim2015physically}
H.~Lim, C.~C. Battaile, J.~D. Carroll, B.~L. Boyce, C.~R. Weinberger, A physically based model of temperature and strain rate dependent yield in {BCC} metals: Implementation into crystal plasticity, Journal of the Mechanics and Physics of Solids 74 (2015) 80--96.

\bibitem{zotov2021molecular}
N.~Zotov, B.~Grabowski, Molecular dynamics simulations of screw dislocation mobility in {BCC} {Nb}, Modelling and Simulation in Materials Science and Engineering 29~(8) (2021) 085007.

\bibitem{suzuki1995plastic}
T.~Suzuki, H.~Koizumi, H.~O. Kirchner, Plastic flow stress of bcc transition metals and the peierls potential, Acta Metallurgica et Materialia 43~(6) (1995) 2177--2187.

\bibitem{proville2012quantum}
L.~Proville, D.~Rodney, M.-C. Marinica, Quantum effect on thermally activated glide of dislocations, Nature Materials 11~(10) (2012) 845--849.

\bibitem{maresca2020theory}
F.~Maresca, W.~A. Curtin, Theory of screw dislocation strengthening in random {BCC} alloys from dilute to “high-entropy” alloys, Acta Materialia 182 (2020) 144--162.

\bibitem{koehler1970attempt}
J.~Koehler, Attempt to design a strong solid, Physical Review B 2~(2) (1970) 547.

\bibitem{rao2000atomistic}
S.~Rao, P.~Hazzledine, Atomistic simulations of dislocation--interface interactions in the {Cu}-{Ni} multilayer system, Philosophical Magazine A 80~(9) (2000) 2011--2040.

\bibitem{bacon1973effect}
D.~Bacon, U.~Kocks, R.~Scattergood, The effect of dislocation self-interaction on the orowan stress, Philosophical Magazine 28~(6) (1973) 1241--1263.

\bibitem{clouet2012screw}
E.~Clouet, Screw dislocation in zirconium: An ab initio study, Physical Review B 86~(14) (2012) 144104.

\bibitem{hirth1966elastic}
J.~Hirth, J.~Lothe, Elastic and core anisotropies for the< 111> screw dislocation in cubic crystals, Physica Status Solidi (b) 15~(2) (1966) 487--494.

\bibitem{fisher1964single}
E.~Fisher, C.~Renken, Single-crystal elastic moduli and the hcp→ bcc transformation in {Ti}, {Zr}, and {Hf}, Physical Review 135~(2A) (1964) A482.

\bibitem{reid1973elastic}
C.~Reid, J.~Routbort, R.~Maynard, Elastic constants of {Ti}--40 at.\% {Nb} at 298 k, Journal of Applied Physics 44~(3) (1973) 1398--1399.

\bibitem{friak2012theory}
M.~Fri{\'a}k, W.~A. Counts, D.~Ma, B.~Sander, D.~Holec, D.~Raabe, J.~Neugebauer, Theory-guided materials design of multi-phase {Ti-Nb} alloys with bone-matching elastic properties, Materials 5~(10) (2012) 1853--1872.

\bibitem{hirth1983theory}
J.~P. Hirth, J.~Lothe, T.~Mura, Theory of dislocations, Journal of Applied Mechanics 50~(2) (1983) 476.

\bibitem{furuhara1995atomic}
T.~Furuhara, T.~Ogawa, T.~Maki, Atomic structure of interphase boundary of an a precipitate plate in a $\beta$ {Ti} [sbnd] {Cr} alloy, Philosophical Magazine Letters 72~(3) (1995) 175--183.

\bibitem{shen2021mechanistic}
X.~Shen, B.~Yao, Z.~Liu, D.~Legut, H.~Zhang, R.~Zhang, Mechanistic insights into interface-facilitated dislocation nucleation and phase transformation at semicoherent bimetal interfaces, International Journal of Plasticity 146 (2021) 103105.

\bibitem{lin2020dislocation}
B.~Lin, J.~Li, Z.~Wang, J.~Wang, Dislocation nucleation from {Zr}--{Nb} bimetal interfaces cooperating with the dynamic evolution of interfacial dislocations, International Journal of Plasticity 135 (2020) 102830.

\bibitem{zhang2021structures}
J.-Y. Zhang, F.-Z. Dai, Z.-P. Sun, W.-Z. Zhang, Structures and energetics of semicoherent interfaces of precipitates in hcp/bcc systems: a molecular dynamics study, Journal of Materials Science \& Technology 67 (2021) 50--60.

\bibitem{wang2013characterizing}
J.~Wang, R.~Zhang, C.~Zhou, I.~J. Beyerlein, A.~Misra, Characterizing interface dislocations by atomically informed frank-bilby theory, Journal of Materials Research 28 (2013) 1646--1657.

\bibitem{wang2014interface}
J.~Wang, R.~Zhang, C.~Zhou, I.~J. Beyerlein, A.~Misra, Interface dislocation patterns and dislocation nucleation in face-centered-cubic and body-centered-cubic bicrystal interfaces, International Journal of Plasticity 53 (2014) 40--55.

\bibitem{hirth2013interface}
J.~Hirth, R.~Pond, R.~Hoagland, X.-Y. Liu, J.~Wang, Interface defects, reference spaces and the frank--bilby equation, Progress in Materials Science 58~(5) (2013) 749--823.

\bibitem{larsen2016robust}
P.~M. Larsen, S.~Schmidt, J.~Schi{\o}tz, Robust structural identification via polyhedral template matching, Modelling and Simulation in Materials Science and Engineering 24~(5) (2016) 055007.

\end{thebibliography}



\section*{Supplementary material}

# Supplementary materials for Interaction of $\langle a \rangle$ prismatic screw dislocations with the $\alpha - \beta$ interface side face in $\alpha - \beta$ Ti alloys

Ali Rida*, Satish I. Rao, Jaafar A. El-Awady*

*Department of Mechanical Engineering, Whiting School of Engineering, Johns Hopkins University, Baltimore, MD 21218, United States*

### S1. The core structure of a $1/2\langle 111 \rangle$ screw dislocation in pure Nb and in $Ti_{60}Nb_{40}$

To determine the stable core structure of the $1/2\langle 111 \rangle$ screw dislocation in the $\beta$ phase (i.e., the $Ti_{60}Nb_{40}$ alloy), a rectangular simulation cell with dimensions $40\sqrt{3}a_\beta \times 34\sqrt{6}a_\beta \times 40\sqrt{2}a_\beta$ (i.e., a total of 651,280 atoms) was constructed. Periodic boundary conditions were employed along the $\mathbf{x} = [111]$ direction (i.e., the dislocation line direction), and free surfaces boundary conditions along the $\mathbf{y} = [\bar{1}\bar{1}2]$ and $\mathbf{z} = [1\bar{1}0]$ directions. A conjugate gradient minimization process was then applied to relax all the components of the pressure tensor and ensure that the free surfaces have zero pressure. Ten different randomly distributed $Ti_{60}Nb_{40}$ configurations were constructed to ensure a correct statistical description of the atomic core structure. A $1/2\langle 111 \rangle$ screw dislocation was then introduced at the center of the simulation cell using its anisotropic elasticity displacement field with its line direction along the periodic $\mathbf{x}$-direction [1]. A conjugate gradient minimization was then applied on the initial atomic positions to determine the relaxed core structures.

The same minimization procedure was also applied to determine the core structure of the $1/2\langle 111 \rangle$ screw dislocation in pure BCC Nb. As before, the screw dislocation is introduced in the simulation cell using its anisotropic elasticity displacement field [1]. However, ten different elastic centers for the initial anisotropic elasticity displacement field were considered to examine the different possible relaxed core structures.

---

*Corresponding authors
  *Email addresses:* `arida1@jhu.edu` (Ali Rida ), `jelawady@jhu.edu` (Jaafar A. El-Awady )

The core structure of a 1/2[111] screw dislocation in pure BCC Nb and in the $Ti_{60}Nb_{40}$ BCC random alloy are shown in Figure S1. The core structure is determined using both the 2D differential displacement plots and the dislocation extraction algorithm (DXA) [2]. In pure Nb, the screw dislocation core is compact, symmetric in the (111) plane, and perfectly straight along the dislocation line, which is similar to other pure BCC metals (e.g. Mo [3], Fe [4], etc.). Interestingly, independent of the initial elastic center of the dislocation in ten different simulations, the equilibrium core configuration was always the compact core shown in Figure S1(a). This indicates that the Ehemann potential does not stabilize any artificial metastable split or hard core at zero applied stress, in agreement with DFT calculations [5]. On the other hand, for the $Ti_{60}Nb_{40}$ alloy the core structure is not compact and it shows kinks spreading at different {110} planes along the dislocation line, as shown in Figure S1(b)-(d). In this case, the screw dislocation core is relaxing to more favorable composition regions to reduce the solute-dislocation interaction energy. Thus, the variations in the core structure along the dislocation line can be viewed as a balance between the dislocation line energy, which increase with the formation of kinks on different {110} planes, and the solute-dislocation interaction energy. Similar core structures were also observed in first principle and MD simulations of other BCC multi-principal element alloys (MPEAs) (*cf.* [6, 7]).

---

**Supplementary References**

[1] M. Yoo, B. Loh, Numerical calculation of elastic properties for straight dislocations in anisotropic crystals., Tech. rep., Oak Ridge National Lab., Tenn. (1971).

[2] A. Stukowski, V. V. Bulatov, A. Arsenlis, Automated identification and indexing of dislocations in crystal interfaces, Modelling and Simulation in Materials Science and Engineering 20 (8) (2012) 085007.

[3] L. Dezerald, D. Rodney, E. Clouet, L. Ventelon, F. Willaime, Plastic anisotropy and dislocation trajectory in BCC metals, Nature Communications 7 (1) (2016) 11695.

[4] L. Ventelon, F. Willaime, E. Clouet, D. Rodney, Ab initio investigation of the peierls potential of screw dislocations in BCC Fe and W, Acta Materialia 61 (11) (2013) 3973–3985.

[5] C. R. Weinberger, G. J. Tucker, S. M. Foiles, Peierls potential of screw dislocations in bcc

transition metals: Predictions from density functional theory, Physical Review B 87 (5) (2013) 054114.

[6] S. Rao, B. Akdim, E. Antillon, C. Woodward, T. Parthasarathy, O. Senkov, Modeling solution hardening in BCC refractory complex concentrated alloys: NbTiZr, Nb1. 5TiZr0. 5 and Nb0. 5TiZr1. 5, Acta Materialia 168 (2019) 222–236.

[7] B. Akdim, C. Woodward, S. Rao, E. Antillon, Predicting core structure variations and spontaneous partial kink formation for 1/2< 111> screw dislocations in three BCC NbTiZr alloys, Scripta Materialia 199 (2021) 113834.

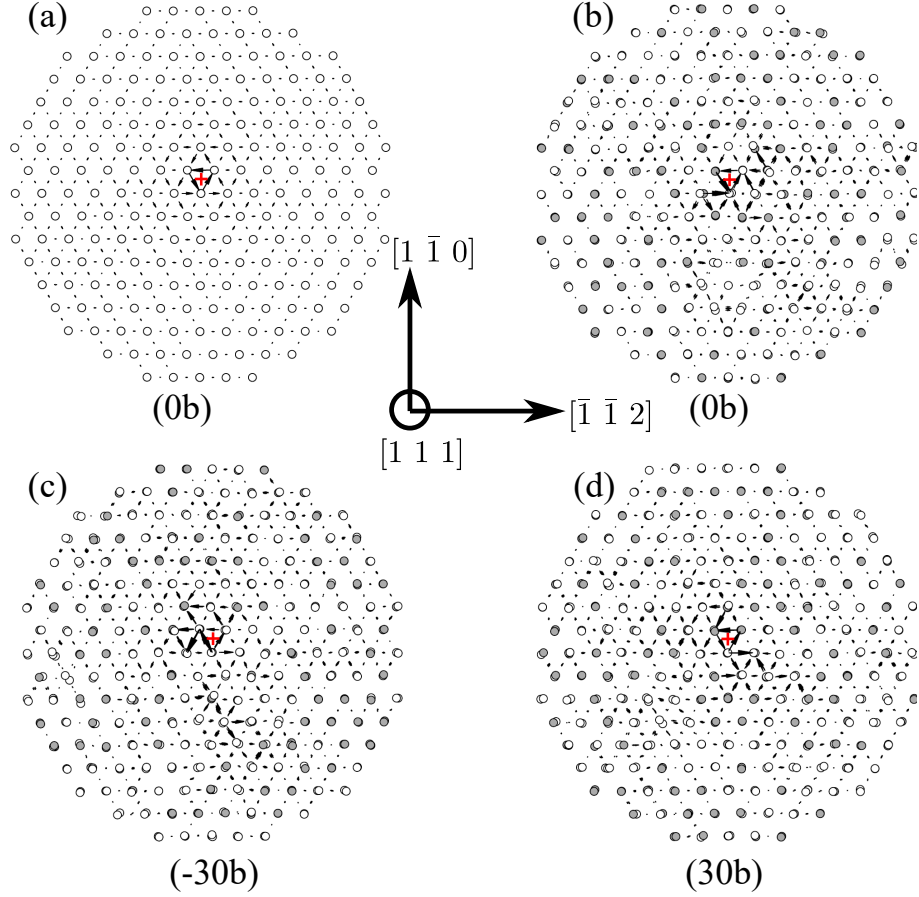

Figure S1: The differential displacement plots of the core structure of the $1/2[111]$ screw dislocation as predicted using the Ehemann potential for: (a) pure Nb, and (b), (c), and (d) $Ti_{60}Nb_{40}$ at different positions along the dislocation line. For each case a '2b' thin slab section perpendicular to the dislocation line is used to generate the differential displacement plots, where $b$ is the Burgers vector magnitude. Open and gray circles indicate the (111) projection of Ti and Nb atoms, respectively. A significant core variation is clearly observed along the dislocation line for the $Ti_{60}Nb_{40}$ case. The position of the screw dislocation center is marked by a red cross as was identified by DXA method in OVITO [2].

4

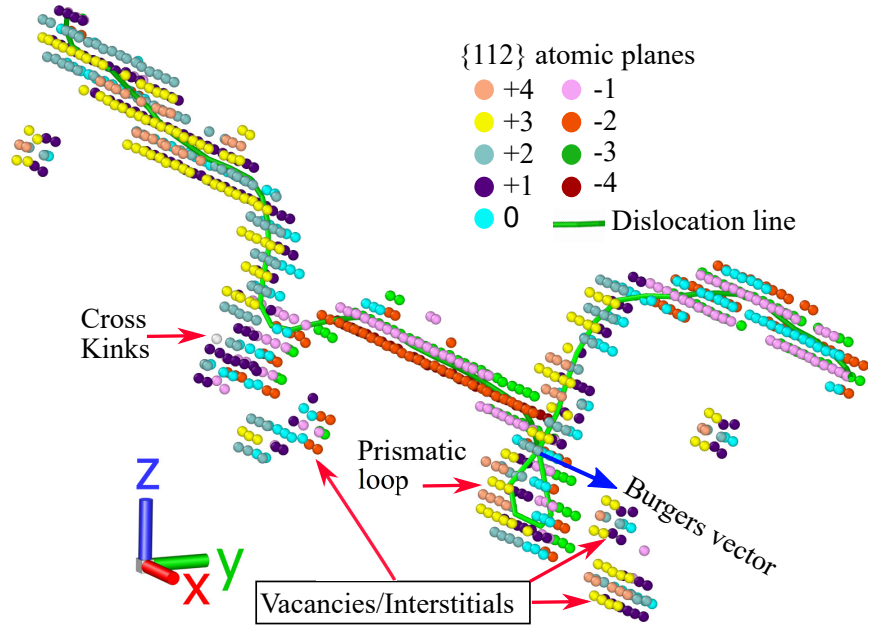

Figure S2: Rough motion and debris production from the 1/2[111] screw dislocation in $Ti_{60}Nb_{40}$ random alloy. Disordered atoms in the 1/2[111] screw dislocation core are shown. Colors indicate the height of atoms relative to the average {112} glide plane. Numbers in color scale indicate the order of atomic planes below and above the average {112} plane. Coloring demonstrates that cross-kinks formation is associated with frequent collisions of kinks at different {110} planes. Prismatic loop resulting from cross-kinks pin the dislocation line thereby contributing to strengthening.

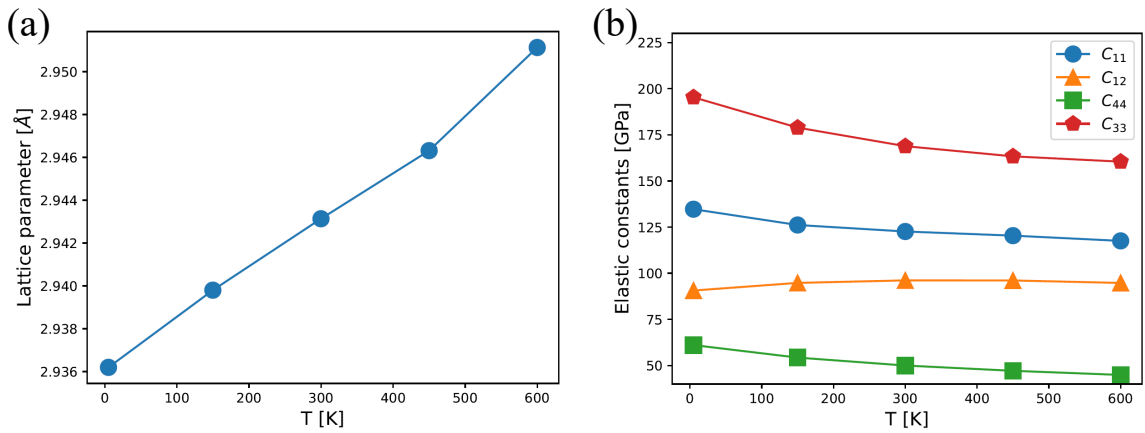

Figure S3: Variation of: (a) the HCP lattice parameter, and (b) the elastic constants, for pure Ti as function of temperature using the Ehemann potential.

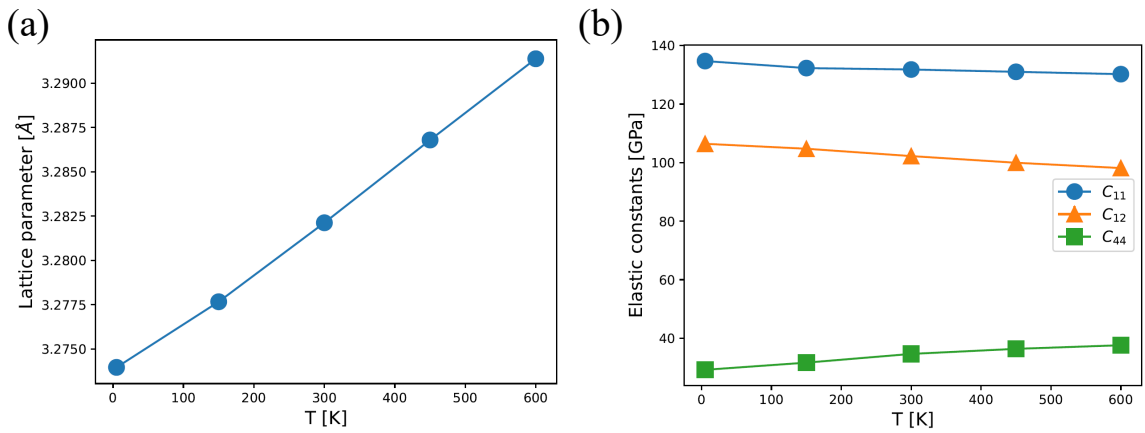

Figure S4: Variation of: (a) the BCC lattice parameter, and (b) the elastic constants for $Ti_{60}$ $Nb_{40}$ random alloy as function of temperature using the Ehemann potential. These calculations are averaged over five different random distributions.

6


\end{document}